\documentclass[11pt,longbibliography,reqno]{amsart}

\usepackage{fullpage}

\usepackage{amsmath}
\usepackage{amsthm}
\usepackage{amssymb}
\usepackage{amsfonts}
\usepackage{mathtools}
\usepackage{paralist}
\usepackage[colorlinks=true,linkcolor=blue,citecolor=blue,urlcolor=blue,breaklinks]{hyperref}
\usepackage{url}
\usepackage{xcolor}
\usepackage{cases}
\usepackage{subfig}
\usepackage{graphicx,epstopdf}
\usepackage{hyphenat}

\allowdisplaybreaks
\mathtoolsset{showonlyrefs}

\numberwithin{equation}{section}

\theoremstyle{plain}
\newtheorem{thm}{Theorem}[section]

\theoremstyle{definition}

\theoremstyle{remark}
\newtheorem{rem}[thm]{Remark}

\theoremstyle{example}

\newcommand{\be}{\begin{equation}}
\newcommand{\ee}{\end{equation}}
\newcommand{\bfig}{\begin{figure}}
\newcommand{\efig}{\end{figure}}
\newcommand{\bt}{\begin{table}}
\newcommand{\et}{\end{table}}
\newcommand{\bc}{\begin{center}}
\newcommand{\ec}{\end{center}}
\newcommand{\ba}{\begin{array}}
\newcommand{\ea}{\end{array}}
\newcommand{\bes}{\begin{equation*}}
\newcommand{\ees}{\end{equation*}}

\renewcommand{\c}{\textrm{max}}

\newcommand{\mt}[1]{\mathrm{#1}}

\def\R{\mathbb{R}}
\def\N{\mathbb{N}}

\def\d{\,\mt{d}}

\def\:{\colon}
\def\e{\varepsilon}

\newcommand{\rmax}{r_{\mt{max}}}
\newcommand{\de}{\Delta\eta}
\newcommand{\dr}{\Delta r}
\newcommand{\tz}{\tilde{z}}

\begin{document}

\title{A revisited Johnson-Mehl-Avrami-Kolmogorov model and the evolution of grain-size distributions in steel}

\author{D. H\"omberg}
\address{Weierstrass Institute for Applied Analysis and Stochastics, Mohrenstr. 39, 10117 Berlin, Germany and Department of Mathematical Sciences, NTNU, Alfred Getz vei 1, 7491 Trondheim, Norway.}
\email{hoemberg@wias-berlin.de}
\author{F. S. Patacchini}
\address{Department of Mathematics, Imperial College London, South Kensington Campus, London SW7 2AZ, UK.}
\email{fsp13@imperial.ac.uk}
\author{K. Sakamoto}
\address{Mathematical Science \& Technology Research Lab., Advanced Technology Research Laboratories, Technical Development Bureau, Nippon Steel \& Sumitomo Metal Corporation, 20-1 Shintomi, Futtsu-City, Chiba, 293-8511, Japan.}
\email{sakamoto.a2c.kenichi@jp.nssmc.com}
\author{J. Zimmer}
\address{Department of Mathematical Sciences, University of Bath, Bath BA2 7AY, UK.}
\email{zimmer@maths.bath.ac.uk}

\date{28 April 2017}

\maketitle

\begin{abstract}
The classical Johnson-Mehl-Avrami-Kolmogorov approach for nucleation and growth models of diffusive phase transitions
  is revisited and applied to model the growth of ferrite in multiphase steels. For the prediction of mechanical
  properties of such steels, a deeper knowledge of the grain structure is essential. To this end, a Fokker-Planck
  evolution law for the volume distribution of ferrite grains is developed and shown to exhibit a log-normally
  distributed solution. Numerical parameter studies are given and confirm expected properties qualitatively. As a
  preparation for future work on parameter identification, a strategy is presented for the comparison of volume
  distributions with area distributions experimentally gained from polished micrograph sections.\\
  
\\
2010 Math Subject Classification: 35Q84, 35Q74, 35K10, 74H40.
\end{abstract}


\section{Introduction}
Steel is still the most important construction material in industrialised countries. Driven especially by the goal of
vehicle weight reduction in automotive industry, the last two decades have seen the development of many new steel
grades, such as dual, trip, or twip steels combining both strength and ductility. The production of these new steels
requires a precise process guidance~\cite{piyada1,piyada2}. It has turned out that best results are achieved if in
addition to temperature measurements, the resulting microstructure is also monitored. Macroscopic phase transition
models allowing for a coupling with finite element simulation of temperature evolution have thus gained increasing
interest.

A classical model to describe diffusive nucleation and growth is the Johnson-Mehl-Avrami\-Kolmogorov (JMAK)
model, developed independently by~\cite{JM39}, \cite{Avrami1,Avrami2,Avrami3}, and~\cite{Kolmogorov}. A review of the
JMAK model can be found in~\cite{FT98}. Assuming constant nucleation and growth rates $\alpha$ and $\rho$,
respectively, it states that the volume fraction $P(t)$ at some time $t$ of a new phase growing from a parent phase by
a nucleation and growth process is given by
\begin{equation}
  \label{eq:JMAK0}
  P(t)=1-\exp\left(-\frac{\pi\alpha\rho^3t^4}{3}\right).
\end{equation}
This JMAK equation is widely used in engineering literature due to its simplicity; in fact, many extensions of it to
situations with non-constant nucleation and growth rates can be found; see for example~\cite{Agarwal1981}.

The first contribution of this paper is to revisit the classical nucleation and growth modelling approach of~\cite{JM39}, \cite{Avrami1,Avrami2,Avrami3}, and~\cite{Kolmogorov}. Specifically, in Section~\ref{sec:avrami} we focus on the growth of ferrite phase from the high
temperature austenite phase, which plays an important role, e.g., in dual phase and trip steels. Ferrite is a solid
solution of carbon in face-centred cubic (f.c.c.) iron. Its time-dependent growth is governed by the diffusion of carbon into the
remaining austenite, thereby enriching its carbon content. The transition naturally ceases when the equilibrium
fraction of carbon in austenite is reached---this is the so-called \emph{soft impingement}. The JMAK model
  employed in this article has indeed been found to be in good agreement with soft impingement, and also with non-random
  nucleation effects~\cite{Bruna2006a}.

An important aim of material simulation is the prediction of mechanical properties. However, especially in
heterogeneous materials such as metals, a representative volume (or concentration) approach is not sufficient to
predict material properties if it does not account for the distribution of grain (or nucleus) sizes. Furthermore, a
macroscopic nucleation and growth model is not capable of resolving mesoscopic grain boundaries. The second
contribution of this paper is to gather additional information about the material heterogeneity by studying the
grain-size distribution. As shown in~\cite{BFB07}, this can then be used by a stochastic homogenisation approach to
derive mechanical properties.

As a conserved quantity, the grain-size (volume) distribution is governed by a Fokker-Planck equation derived and solved in Section~\ref{sec:grain-size}. Grain-size distributions in austenite and ferrite are of log-normal type~\cite{MGHM96}, as is the solution to the Fokker-Planck equation we study, up to a convolution with the initial profile. In fact, we rigorously obtain a log-normal solution for all times only if the initial profile is itself log-normal. Nevertheless, for any admissible inital datum the solution profile we find is log-normal asymptotically in time in numerical tests. For a different application of similar
Fokker-Planck models; see for example~\cite{CPP09}. One core aspect of our Fokker-Planck model is that it naturally couples the macroscopic scale (ferrite phase fraction) with the mesoscopic scale (ferrite grain-size distribution) via a first-moment constraint given in~\eqref{eq:moment-constraint}. In Section~\ref{sec:num-study} we present simulations and a
numerical parameter study. In a forthcoming paper, we will discuss parameter identification issues for this
Fokker-Planck model and compare it to real-world data. As a preparation, we discuss in Section~\ref{sec:model-data}
how volume distributions can be compared with area distributions drawn from polished micrograph sections using an approach by~\cite{Huber}. For this question we also refer
  the reader to~\cite{Takayama1991a}. For results
about the identification of temperature-dependent growth rates exploiting dilatometer experiments, we refer
to~\cite{HTY09,HLSY13}.

The promising feature of this approach is that it allows for an easy calculation of grain-size distributions from a macroscopic level without explicit mesoscopic simulations as in the phase-field approach, opening up at least two interesting and obvious areas of further research. The first one is the inclusion of thermal effects by a spatial two-scale model combining area space with the macroscopic specimen space. The second one is the use of these grain-size distributions for a computation of homogenised mechanical properties.

Somewhat similarly to our approach, the evolution of the absolute number of grains has also been modelled by~\cite{Teran2010a} and~\cite{Bergmann2008a} using a continuity equation of conservative type with a reaction term. There too, the authors obtain an analytic expression for the solution which, remarkably, is asymptotically (as $t$ goes to infinity) log-normally distributed in one space dimension. In higher space dimensions they obtain solutions which are not log-normal, but still qualitatively close to it. In~\cite{Teran2010a}, the authors link their model to the JMAK model by choosing a specific reaction term and specific nucleation and growth rates. In contrast, in the present paper we make this link through the very definition of the first moment of the grain-size distribution. We also remark that our approach can easily be adapted to arbitrary space dimensions and that obtaining a log-normally distributed explicit solution does not depend on the choice of this dimension; indeed, the JMAK model can be extended to any dimension and then linked to our Fokker-Planck equation which is dimension-free. Let us mention as well that~\cite{Tomellini2003a} uses a JMAK approach similar to the one of Section~\ref{sec:avrami} and, using Fick's law, couples it to a diffusion equation for the new phase concentration.

It is noteworthy at this point that our approach should not be confused with grain boundary character distribution
evolution models of Fokker-Planck type as they have been investigated in a series of papers
by~\cite{Barmak2011b,BEEEKST12}; see also the references therein. While these authors study \emph{coarsening effects}
in polycrystalline solids, i.e., a single phase situation, the present paper is concerned with the evolution of the
grain-size distribution during an irreversible phase transition without coarsening. Here, no grain can grow at the
expense of others, no grain shrinks, and no grain can grow into others when touching---this is the so-called \emph{hard
  impingement}. Similarly, the approach taken in this paper for nucleation and growth processes is different from the
Becker-D\"oring type models of coagulating particles or droplets; see, e.g.,~\cite{Ball1986a} and~\cite{Penrose1989a}.

\section{The revisited JMAK model}
\label{sec:avrami}

Consider a bounded domain $\Omega \subset \R^3$, whose volume is denoted by $V=|\Omega|$, composed exclusively of an
austenite phase and a ferrite phase, and where austenite may transform into ferrite as time increases. The sub-volume
of austenite present at time $t$ is denoted by $V_\mt{A}(t)$, and the one of ferrite by $V_\mt{F}(t)$. By conservation
of volume we have $V = V_\mt{A}(t)+V_\mt{F}(t)$ for all $t \in [0,T]$, where $T > 0$ is a fixed final time. Then, the
volume phase fraction of ferrite is defined by
\begin{equation}
  \label{eq:P}
  P(t) = \frac{V_\mt{F}(t)}{V}.
\end{equation}
To derive our model, we assume that the phase transformation happens isothermally at temperature $\theta >0$, although
this can be easily generalised.

We assume that ferrite grains appear randomly in the austenite matrix $\Omega$ with nucleation rate
$\alpha = \alpha(\theta)$ (number of grains per unit time per unit volume) and grow isotropically, that is, as spheres,
with growth rate $\rho(t,\theta) = \rho(t)$ (length per unit time). We suppose that when two growing grains touch,
these cannot grow into each other and thus only continue growing towards the ``free'' directions, which is the hard
impingement assumption. After two grains meet, they therefore do not look as spheres anymore, but rather as the union
of two intersected spheres. Let us point out that hard impingement also describes the ``interaction'' between the
grains and the boundary of the domain $\Omega$ when these touch. In this setting, the volume occupied at time $t$ by an
isolated ferrite grain born at time $\tau$ is
\begin{equation} 
  \label{single-vol}
  \nu(t,\tau) = \frac{4\pi}{3} \left(\int_\tau^t \rho(s)\d s\right )^3.
\end{equation}
Consider an extended volume, denoted by $V^\mt{ext}(t)$, which is the total volume occupied by all ferrite grains at
time $t$, assuming temporarily that they may grow into each other. This gives, using~\eqref{eq:P},
\begin{equation}
  V_\mt{F}^\mt{ext}(t)= V \alpha \int_0^t \nu(t,\tau) \d\tau.
\end{equation}
Invoking the Avrami correction (see~\cite{Avrami1,Avrami2,Avrami3} for a derivation, and also~\cite{Kolmogorov}) to
incorporate hard impingement, we have
\begin{equation}
  V\mt{d} P(t) = \mt{d}V_\mt{F}(t) = \left(1 - P(t)\right)\mt{d}V_\mt{F}^\mt{ext}(t).
\end{equation}
We remark that the Avrami correction is only an approximation, due to possible overgrowth of phantom
  nuclei~\cite{Tomellini1997a}.  By integrating the above equation, using~\eqref{single-vol} and supposing that
$P(0) = 0$, we get
\begin{equation}
  \label{eq:logP}
  -\log \left(1-P(t)\right) = \frac{4\pi\alpha}{3}\int_0^t \left(\int_\tau^t \rho(s) \d s\right )^3 \d\tau.
\end{equation}
Note that, by assuming that $\rho(t) = \rho$ does not depend on time, and by taking the exponential of both sides
of~\eqref{eq:logP}, we recover~\eqref{eq:JMAK0}. Equation~\eqref{eq:logP} yields 
\begin{equation}
  \label{eq:diff}
  \begin{cases}
    P'(t) =  4\pi\alpha\rho(t)(1-P(t))\displaystyle \int_0^t \left(\int_\tau^t \rho(s) \d s\right)^2 \d\tau,\\
    P(0) = 0.
  \end{cases}
\end{equation}
In order to close the differential equation~\eqref{eq:diff}, we need now to choose a law for the evolution of the
growth rate $\rho$. This is where we incorporate soft impingement into the model. This means that the transformation
ceases naturally when the actual carbon concentration in austenite, $C_\mt{A}(t)$, reaches the equilibrium value
$C_\mt{A}^\mt{eq} = C_\mt{A}^\mt{eq}(\theta)$, corresponding to an equilibrium volume
$V_\mt{F}^\mt{eq} = V_\mt{F}^\mt{eq}(\theta)$ and equilibrium fraction $P^\mt{eq} = V_\mt{F}^\mt{eq}/V$. Then
$C_\mt{A}(t)$ can be computed from mass conservation by assuming that the carbon concentration in ferrite is constant
and equal to its equilibrium value $C_\mt{F}^\mt{eq} = C_\mt{F}^\mt{eq}(\theta)$ (defined as the carbon concentration in
ferrite when $C_\mt{A}^\mt{eq}$ is reached), i.e.,
\begin{equation}
  C = C_\mt{F}^\mt{eq} P(t) +C_\mt{A}(t)(1 - P(t)),
\end{equation}
where $C$ is the overall carbon concentration in the steel sample $\Omega$. From this it follows that, if
$C_\mt{A}^\mt{eq}\neq C_\mt{F}^\mt{eq}$ (otherwise nothing happens), the equilibrium volume fraction of ferrite
satisfies
\begin{equation}
  P^\mt{eq} = \frac{C_\mt{A}^\mt{eq} - C}{C_\mt{A}^\mt{eq}-C_\mt{F}^\mt{eq}} \quad \mbox{and} 
  \quad \frac{P^\mt{eq}-P(t)}{1-P(t)} = \frac{C_\mt{A}^\mt{eq} - C_\mt{A}(t)}{C_\mt{A}^\mt{eq} - C_\mt{F}^\mt{eq}}, 
\end{equation}
so that the ratio $(P^\mt{eq}-P(t))/(1-P(t))$ equals the \emph{supersaturation}. We then require the growth rate
$\rho(t)$ to be proportional to $C_\mt{A}^\mt{eq}-C_\mt{A}(t)$, and we make the choice
\begin{equation}\label{eq:growth-rate}
  \rho(t) = \frac{\rho_*}{ct^\gamma} \frac{P^\mt{eq} - P(t)}{1-P(t)}, \quad 0 \leq \gamma < 1,
\end{equation}
where $\rho_* = \rho_*(\theta)>0$ is some reference growth rate, and $c$ is a constant with the same dimension as
$t^{-\gamma}$; for simplicity, we take $c:=1$. The term $t^\gamma$ allows for the description of time-dependent growth
rates, independently of soft impingement. In the case of classical diffusional growth, we have $\gamma=0.5$. This
ansatz for the growth rate then results in the integro-differential equation model
\begin{equation}
  P'(t) = 4\pi\alpha\rho_*t^{-\gamma}(P^\mt{eq}-P(t)) \int_0^t \left( \int_\tau^t \frac{\rho_*}{s^\gamma} \frac{P^\mt{eq} 
      - P(s)}{1-P(s)} \d s \right)^2 \d\tau.
\end{equation}
Note that the equilibrium value $P^\mt{eq}$ is only reached asymptotically. This equation can be dealt with by
transformation to a system of ODEs. To this end, we perform the substitutions
\begin{equation}\label{eq:ode-jmak}
  z(t,\tau) =  \int_\tau^t \rho(s) \d s,\quad y(t) = \alpha\int_0^t z(t,\tau)^2 \d\tau,\quad x(t) 
  = \alpha\int_0^t z(t,\tau) \d\tau,\quad w(t) = \alpha t.
\end{equation}
Altogether we obtain
\begin{equation}
  \begin{cases}
    w'(t) = \alpha,\quad x'(t) = \rho(t) w(t), \quad y'(t) = 2\rho(t) x(t),\\ 
    P'(t) = 4\pi (1-P(t)) \rho(t) y(t),
  \end{cases}
\end{equation}
with $w(0)=x(0)=y(0)=P(0)=0$. We can finally introduce the number of grains born until time $t$ per unit volume
\begin{equation}
  \label{eq:N}
  N(t) = \alpha\int_0^t \left(1-\frac{P(\tau)}{P^\mt{eq}}\right) \d \tau.
\end{equation} 
The expression~\eqref{eq:N} takes soft impingement into account as well by requiring that nucleation stops when $P^\mt{eq}$ is
reached.

\section{The grain-size distribution model}
\label{sec:grain-size}

\subsection{Derivation of the governing equation}
The \emph{volume distribution} of ferrite grains
\begin{equation}
  \label{eq:phi}
  \phi(\nu,t) \: (0,\infty)\times [t_0,T] \to [0,\infty)
\end{equation}
counts, at time $t$, the number of grains of volume $\nu$ per unit volume, normalised by the total number of
grains. This means that, for any $\nu_2 > \nu_1 \geq 0$, the quantity $\int_{\nu_1}^{\nu_2} \phi(\nu,t) \d \nu$ is the
relative number of grains with volumes in $[\nu_1,\nu_2]$, which implies that $\phi(\cdot,t)$ is a probability density
on $(0,\infty)$, that is, $\int_0^\infty \phi(\nu,t) \d \nu = 1$. In~\eqref{eq:phi}, $t_0>0$ is a small
\emph{incubation} time before which the notion of volume distribution does not make physical sense. Indeed, there exists a small critical
  average grain volume below which we are unable to describe physically, or simply observe, the evolution of the ferrite grain-size distribution in
  the specimen. The strictly positive incubation time $t_0$ is defined as being the smallest time after which the average ferrite
  volume in the specimen has reached this critical volume (see~\cite{THhandbook} for an account on the notion of incubation time). Therefore, while the JMAK model starts at $t=0$ with zero volume fraction of ferrite as in~\eqref{eq:diff}---which would correspond to a volume distribution which is a Dirac mass at the origin---the ferrite volume distribution model that we derive below is only meaningful for times $t\geq t_0$.
  
Since $\phi(\cdot,t)$ has conserved unit mass over all times $t \in [t_0,T]$, we assume that $\phi$ satisfies the
continuity equation $\phi_t + J_\nu = 0$, with
\begin{equation}
  J(\nu,t) = \mu_1(\nu,t) \phi - (\mu_2(\nu,t)\phi)_\nu,
\end{equation}
where the mobility terms $\mu_1$ and $\mu_2$ are assumed to be separable, $\mu_1(\nu,t) = \mu_{11}(t)\mu_{12}(\nu)$ and
$\mu_2(\nu,t) = \mu_{21}(t)\mu_{22}(\nu)$. Here, we suppose that $\mu_{12}(\nu) = \nu$ and $\mu_{22}(\nu) =
\nu^2$.
These choices are justified \emph{a posteriori}: they allow the derivation of an explicit solution for the volume
distribution which is, up to a convolution with the initial datum, log-normally distributed (see Section~\ref{subsec:formula}); and this log-normal behaviour is
experimentally observed. Also, we write $u(t):=\mu_{11}(t)$ and $\beta(t):=\mu_{21}(t)$, where we assume
$\beta(t) = f(u(t))$ for some function $f\in C^\infty(\R)$ such that $f(0) = 0$ and $f(u) >0$ for all $u\neq 0$. The requirement that
$f(0) = 0$ is physically justified by the fact that the volume distribution stops evolving as soon as the convection
vanishes, and therefore the diffusion has to vanish as well. The condition $f(u)>0$ for all $u\neq 0$ is needed to avoid backward
diffusion in the case the convection velocity $u$ becomes negative. Indeed, as it becomes clearer in the following,
this may happen when the nucleation rate ``beats'' the grain growth and thus ``drags'' the volume distribution profile
towards the left. In this paper, we choose $f(u) = \beta_1 u^2$, where $\beta_1 > 0$, although in Section~\ref{subsec:blowup}
we show the appearance of infinite-time blow-up if we violate the condition $f(0) = 0$ for the special case
$f(u) = \beta_0 + \beta_1 u^2$ with $\beta_0>0$. All in all, the volume distribution $\phi$ is assumed to satisfy the Fokker-Planck
equation
\begin{equation} 
  \label{eq:fp}
  \begin{cases}
    \phi_t = - u(t) (\nu\phi)_\nu + \beta_1 u(t)^2 (\nu^2\phi)_{\nu\nu},\\
    \phi(\nu,t_0) = \phi_0(\nu),
  \end{cases} \mbox{for all $(\nu,t) \in (0,\infty) \times (t_0,T]$},
\end{equation}
where $\phi_0 \in C^0(0,\infty) \cap L^\infty(0,\infty)$ is a probability density. 
  
An essential feature of the present grain-size distribution model lies in the fact that we can directly link it to the revisited JMAK model developed in Section~\ref{sec:avrami} using the natural moment relation
\begin{equation} 
\label{eq:moment-constraint}
	\int_0^\infty \nu \phi(\nu,t) \d \nu = \frac{P(t)}{N(t)}=:g(t) \quad \mbox{for all $t \in [t_0,T]$},
\end{equation}
where we recall from~\eqref{eq:P} that $P$ is the volume phase fraction of ferrite and from~\eqref{eq:N} that $N$ is the relative number of ferrite grains. The left-hand side of~\eqref{eq:moment-constraint} is the first
moment of $\phi(\cdot,t)$ at time $t$, i.e., the mean volume of the grains. This equation bridges the meso- and
macroscopic scales, giving us another nice feature of the JMAK model, namely that it allows to compute the mean grain
size without relying on further mesoscopic information.  We refer to~\cite{Carlen2003a}, \cite{Tudorascu2011a} and~\cite{Dreyer2015a} for mathematical studies of Fokker-Planck/gradient flow equations with moment constraints. Equation~\eqref{eq:moment-constraint} is a constraint that is imposed by the JMAK model on the Fokker-Planck model~\eqref{eq:fp}; therefore the volume distribution $\phi$ satisfies the coupled system
\begin{equation} 
  \label{eq:fp-coupled}
  \begin{cases}
    \phi_t = - u(t) (\nu\phi)_\nu + \beta_1 u(t)^2 (\nu^2\phi)_{\nu\nu},\\
    \int_0^\infty \nu \phi(\nu,t) \d \nu = g(t),\\
    \phi(\nu,t_0) = \phi_0(\nu),
  \end{cases} \mbox{for all $(\nu,t) \in (0,\infty) \times (t_0,T]$}.
\end{equation}
Constraint~\eqref{eq:moment-constraint} also enforces a relation between $g$ and the convection velocity $u$. Indeed,
\begin{align}
  g'(t) &= \displaystyle \int_0^\infty \nu \phi_t(\nu,t) \d\nu = - u(t) 
            \displaystyle \int_0^\infty \nu (\nu\phi)_\nu(\nu,t) \d \nu
            + \beta_1u(t)^2 \displaystyle \int_0^\infty \nu (\nu^2 \phi)_{\nu\nu}(\nu,t) \d \nu \\
          & = u(t) \displaystyle \int_0^\infty \nu\phi(\nu,t) \d \nu - \beta_1u(t)^2 
            \displaystyle \int_0^\infty (\nu^2\phi)_\nu(\nu,t) \d \nu \\
          & = u(t) \displaystyle \int_0^\infty \nu\phi(\nu,t) \d \nu = u(t)g(t),
\end{align}
where we implicitly need that
\begin{equation} 
  \label{eq:decay-conditions}
  \begin{cases} 
    \displaystyle \lim_{\nu\to 0} \nu^2\phi(\nu,t) = \lim_{\nu\to +\infty} \nu^2\phi(\nu,t) = 0,\\ 
    \displaystyle \lim_{\nu\to 0} \nu^3\phi_\nu(\nu,t) = \lim_{\nu\to \infty} \nu^3\phi_\nu(\nu,t) = 0, 
  \end{cases}
\end{equation}
in order to carry out the integrations by parts. This gives
\begin{equation} 
  \label{eq:v}
  u(t) = \frac{g'(t)}{g(t)} = (\log \circ g)'(t) \quad \mbox{for all $t \in [t_0,T]$},
\end{equation}
or, equivalently,
\begin{equation} 
  \label{eq:moments}
  g(t) = g_0 e^{a(t)} \quad \mbox{for all $t\in[t_0,T]$},
\end{equation}
where $g_0:=g(0)$ and $a(t) := \int_{t_0}^t u(s) \d s$. Equation~\eqref{eq:v} tells us that the mesoscopic convection velocity $u$ (and therefore the diffusion coefficient $\beta$, up to
the multiplicative constant $\beta_1$) is determined by the evolution of the macroscopic quantity $g$ given to us by
the model in Section~\ref{sec:avrami}. Conversely, we can also see the convection velocity $u$ as a measure of the evolution of
$g$. Again, here, we see the coupling between the meso- and macroscopic scales, and the system~\eqref{eq:fp-coupled} is equivalent to
\begin{equation} 
  \label{eq:fp-coupled-2}
  \begin{cases}
    \phi_t = - u(t) (\nu\phi)_\nu + \beta_1 u(t)^2 (\nu^2\phi)_{\nu\nu},\\
    u(t) = (\log \circ g)'(t),\\
    \phi(\nu,t_0) = \phi_0(\nu),
  \end{cases} \mbox{for all $(\nu,t) \in (0,\infty) \times (t_0,T]$}.
\end{equation}

\begin{rem}
As already discussed at the beginning of this section, we cannot hope to describe physically the evolution of the ferrite volume distribution before the incubation time $t_0>0$ is reached. We observe, from a mathematical point of view, that~\eqref{eq:moments} implies that if $t_0$ was taken to be zero, i.e., $\phi_0$ was a Dirac mass at the
origin according to the zero-fraction initial condition in~\eqref{eq:diff}, then any solution to~\eqref{eq:fp} would stay equal to $\phi_0$ for all times, that is, nothing would happen. This
reflects the fact that, even mathematically, our Fokker-Planck model is unable to describe the evolution of $\phi$ for early times.
\end{rem}

\begin{rem}
In~\eqref{eq:fp}, as well as in~\eqref{eq:fp-coupled} and~\eqref{eq:fp-coupled-2}, the volume domain, $(0,\infty)$, is unbounded, which allows grains to grow instantaneously arbitrarily large. Our Fokker-Planck model can therefore be rigorously valid only for unbounded specimens and unbounded grain growth rates, although the JMAK model developed in Section~\ref{sec:avrami} requires the specimen to be a bounded one and the growth rate~\eqref{eq:growth-rate} to be finite. Given that the grains are typically small relative to the size of the specimen, one would expect our model to be a good approximation, which we can control by quantifying the ``portion'' of ferrite volume distribution $\phi$ having a larger volume than a time-dependent maximal volume, $\nu_\c$, imposed by the finiteness of the specimen and the growth rate. Alternatively, we give now a possible improvement of our model taking the boundedness of the specimen and growth rate into account. We simply consider~\eqref{eq:fp} and impose a boundary condition at $\nu_\c$:
\begin{equation} 
\label{eq:fp-cut}
	\begin{cases}
		\phi_t = - u(t) (\nu\phi)_\nu + \beta_1 u(t)^2 (\nu^2\phi)_{\nu\nu}, \\
		\phi(\nu_\c(t),t) = 0,\\
    		\phi(\nu,t_0) = \phi_0(\nu),
\end{cases} \mbox{for all $(\nu,t) \in (0,\nu_\c(t)) \times (t_0,T]$}.
\end{equation}
Obtaining an exact value for this maximal volume $\nu_\c(t)$ may not be possible; nevertheless the JMAK model tells us that it has to satisfy $\nu_\c(t) \leq \min( P^\mt{eq}V, (4\pi/3) z(t,0)^3 )$ for all $t\geq t_0$, where $P^\mt{eq}$ is the equilibrium volume fraction of ferrite, $V$ is the volume of the specimen, and $z$ is as in~\eqref{eq:ode-jmak}. The fact that $\nu_\c(t)$ is bounded by $P^\mt{eq}V$ makes sure that the equilibrium volume (and thus the volume of the specimen) is never exceeded, whereas the fact that $\nu_\c(t)$ is bounded by $(4\pi/3) z(t,0)^3$ ensures that the maximal grain volume allowed by the growth rate at time $t$ is not violated. Unfortunately, there is no explicit solution formula for~\eqref{eq:fp-cut} akin to that for~\eqref{eq:fp} derived in Section~\ref{subsec:formula}. Nevertheless, existence of a solution is known~\cite[Theorem 16.3.1]{Cannon1984a} and we believe that, qualitatively, such a solution is very similar to that of Section~\ref{subsec:formula} for the unbounded case. One way to support this would be to prove that if $\phi_{\nu_\c}$ is solution to~\eqref{eq:fp-cut} and $\phi_{\nu_\c} \to \phi$ (in some sense) as $\min_t (\nu_\c(t)) \to\infty$ for some probability density $\phi$, then $\phi$ must be solution to~\eqref{eq:fp}. Because of the reasons just mentioned, we decide to focus in this paper on~\eqref{eq:fp} only and we leave the study of~\eqref{eq:fp-cut} to a future work. Note nonetheless that in Section~\ref{sec:model-data} we actually derive a relation between volume and area distributions in the case of a bounded specimen; Section~\ref{sec:model-data}, however, is mostly independent of the rest of the paper and is mainly there to motivate a forthcoming paper.
\end{rem}

\begin{rem}
  \label{rem:radius}
  If one assumes that hard impingement in negligible (for example, if the final time is
  very small or the nucleation and growth rates are very small), then grains are exact, non-intersected spheres and
  one may equivalently employ the \emph{radius distribution} $\psi$ in place of the volume distribution $\phi$. The
  radius distribution
  \begin{equation}
    \psi(r,t) \: (0,\infty)\times [t_0,T] \to [0,\infty)
  \end{equation}
  counts, at time $t$, the number of grains of radius $r$ per unit radius, normalised by the total number of grains,
  which leads, as for the volume distribution, to $\psi(\cdot,t)$ being a probability density on $(0,\infty)$. Since
  grains are spheres, there is a direct one-to-one relation between $\phi$ and $\psi$ as the map
  $A\: [0,\infty) \to [0,\infty), r \mapsto 4\pi r^3/3$, is a bijection. Indeed, this implies that, for all
  $r_2 > r_1 \geq 0$,
  \begin{equation}
    \int_{r_1}^{r_2} \psi(r,t) \d r = \int_{A(r_1)}^{A(r_2)} \phi(\nu,t) \d \nu = \int_{r_1}^{r_2} \phi(A(r),t) A'(r) \d r 
    = \int_{r_1}^{r_2} \phi\left(\frac{4\pi r^3}{3},t\right) 4\pi r^2 \d r,
  \end{equation}
  by a simple change of variable $x \to A(r)$. This equality being true for all $r_2 > r_1 \geq 0$, we get
  \begin{equation}
    \textstyle{\psi(r,t) = 4\pi r^2 \phi\left(\frac{4\pi}{3}r^3,t\right) \quad \mbox{for all $r \in (0,\infty)$}.}
  \end{equation}
From the inverse transformation, one gets
\begin{equation}
  \textstyle{\phi(\nu,t) = (4\pi)^{-1/3} (3\nu)^{-2/3} \psi\left(\left(\frac{3\nu}{4\pi}\right)^{1/3},t\right) 
    \quad \mbox{for all $\nu \in (0,\infty)$}.}
\end{equation}
Relation~\eqref{eq:moment-constraint} then becomes
\begin{equation}
  g(t) = \frac{4\pi}{3} \int_0^\infty r^3\psi(r,t) \d r.
\end{equation}
\end{rem}

\subsection{A solution formula for the volume distribution}
\label{subsec:formula}

We now derive an explicit solution for the Fokker-Planck equation~\eqref{eq:fp}, which we later couple to the moment constraint~\eqref{eq:moment-constraint} as in~\eqref{eq:fp-coupled} and~\eqref{eq:fp-coupled-2}. We introduce the transformation of
variables
\begin{align}
	\xi&:= \log(\nu) +  b(t) - a(t), \\
	\tau &:= b(t),\\
	h(\xi,\tau) &:= \nu \phi(\nu,t),
\end{align}
with $h(\xi,\tau) \: \mathbb{R} \times [0,b(T)] \to \left[0,\infty\right)$ and where $a(t) := \int_{t_0}^t u(s) \d s$ and $b(t):= \int_{t_0}^t \beta(s) \d s = \beta_1 \int_{t_0}^t u(s)^2\d s$. The fact that $b$ is increasing allows
us to introduce the time change of variables $\tau = b(t)$; this justifies the requirement that $f(u)>0$ for all $u\neq0$ in
$\beta(t) = f(u(t))$ from a mathematical viewpoint. We see that $h(\xi,\tau) $ is governed by the linear heat equation
\begin{equation}
  \begin{cases}
    h_\tau = h_{\xi\xi}, \\
    h(\xi,0) = \mathrm{e}^{\xi} \phi_0(\mathrm{e}^{\xi}),
  \end{cases} 
  \mbox{for all $(\xi,\tau) \in \mathbb{R} \times (0,b(T)]$}.
\end{equation} 
It is well-known that $h$ is given by
\begin{equation}
  \label{eq:sol}
  h(\xi,\tau) = (K(\cdot,\tau) \ast \Phi_0)(\xi) \quad \mbox{for all $(\xi,\tau) \in \mathbb{R} \times (0,b(T)]$},
\end{equation}
where $\ast$ is the convolution and
\begin{equation}
  K(\xi,\tau) = (4 \pi \tau)^{-1/2} \exp\left(-\xi^2/(4 \tau)\right) \quad \mbox{and} 
  \quad \Phi_0(\xi) = \mathrm{e}^{\xi} \phi_0(\mathrm{e}^{\xi}).
\end{equation}
Transforming back to the original variables $\nu$ and $t$ we finally obtain
\begin{equation} 
  \label{eq:sol1}
  \phi(\nu,t) = \nu^{-1}(K(\cdot,b(t))  \ast \Phi_0)(\log(\nu) + b(t) - a(t)) \quad \mbox{for all 
    $(\nu,t) \in (0,\infty) \times (t_0,T]$}.
\end{equation}
We now see that the resulting solution is, up to a convolution with the initial distribution, log-normal. As already mentioned, this justifies the choice of the mobility terms in the Fokker-Planck equation~\eqref{eq:fp} made earlier, since experiments strongly suggest a log-normal shape for $\phi$~\cite{MGHM96}. We note that, in ~\eqref{eq:sol1}, $\phi(\cdot,t)$ is rigorously log-normally distributed for all times if $\phi_0(\cdot)$ is too. Indeed, in this case $\Phi_0$ is normal and thus the convolution in~\eqref{eq:sol1} is between two normal distributions and is therefore a normal distribution itself, evaluated at $\log(\nu)$, which shows that $\phi(\cdot,t)$ is log-normal for all times. Otherwise, if $\phi_0(\cdot)$ is not log-normally distributed we can only infer from numerical tests that $\phi(\cdot,t)$ converges to a log-normal profile as $t$ increases; see Section~\ref{subsec:initial-profile} where we simulate the evolution of the solution for an initial datum which is not log-normal.

Note that the decay conditions~\eqref{eq:decay-conditions} are satisfied by the solution in~\eqref{eq:sol1} as long as it holds that
$\int_0^\infty \nu^2 \phi_0(\nu) \d\nu<\infty$; see Proposition 3.4 in~\cite{Pa13}. Also, by Propositions 3.1 and
3.2 in~\cite{Pa13}, we have that $\phi$ as given in~\eqref{eq:sol1} satisfies $\phi(\cdot,t) \to \phi_0(\cdot)$ as $t\to t_0$ and $\phi \in C^{\infty,0}((0,\infty)\times[t_0,\infty))$.

\section{Numerical simulations}
\label{sec:num-study}

We study the general qualitative behaviour of the model derived in Section~\ref{sec:avrami} and the
Fokker-Planck equations~\eqref{eq:fp-coupled} and~\eqref{eq:fp-coupled-2}. As initial distribution, except in Section~\ref{subsec:initial-profile}, we take the log-normal profile
\begin{equation}\label{eq:initial}
  \phi_0(\nu) = (\nu\sigma_0\sqrt{2\pi})^{-1}\exp(-(\log(\nu) - \mu_0)^2/(2\sigma_0^2)) \quad \mbox{for all $\nu \in (0,\infty)$},
\end{equation}
with $\mu_0 = \log(g_0) - \sigma_0^2/2$; in fact, $g_0 = \int_0^\infty \nu \phi_0(\nu) \d\nu = \exp(\mu_0 + \sigma_0^2/2)$. The standard variation
$\sigma_0$ cannot be extracted from the model in Section~\ref{sec:avrami} (unlike $g_0$), and is therefore an additional
parameter. Unless mentioned otherwise, the simulations below approximate~\eqref{eq:diff} and plot~\eqref{eq:sol1} for
the parameters
\begin{equation}\label{eq:parameters}
  \rho_* = 1, \quad P^\mt{eq} = 0.45, \quad \alpha = 0.001, \quad \gamma = 0.5, \quad \beta_1 = 0.01, 
  \quad t_0 = 0.3387, \quad \sigma_0 = 0.4.
\end{equation} 
The value of $t_0$ is arbitrary and is only chosen as above for convenience in the following simulations.

\subsection{The main quantities}

From Figures~\ref{fig:P-diffusion} and~\ref{fig:g-diffusion}, we can see that the quantities $g(t)$ and $P(t)$ are
sigmoid functions, reaching ``quickly'' values close to their equilibrium. The evolution of the log-normal volume
distribution of ferrite grains $\phi(\cdot,t)$ is given in Figure~\ref{fig:t150-diffusion}.

\bfig[!ht]
\centerline{\subfloat[Evolution of $P(t)$]{\includegraphics[scale=1.075]{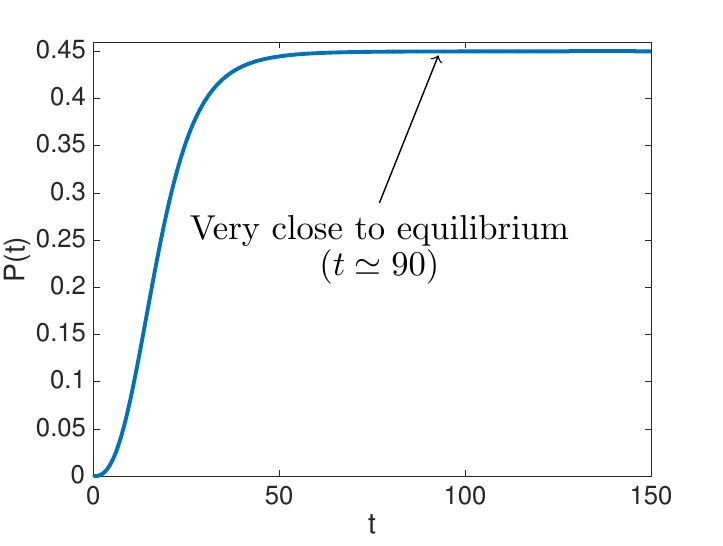}%
\label{fig:P-diffusion}}
\hfil
\subfloat[Evolution of $g(t) := P(t)/N(t) = \int_0^\infty \nu\phi(\nu,t) \d \nu$]{\includegraphics[scale=0.4]{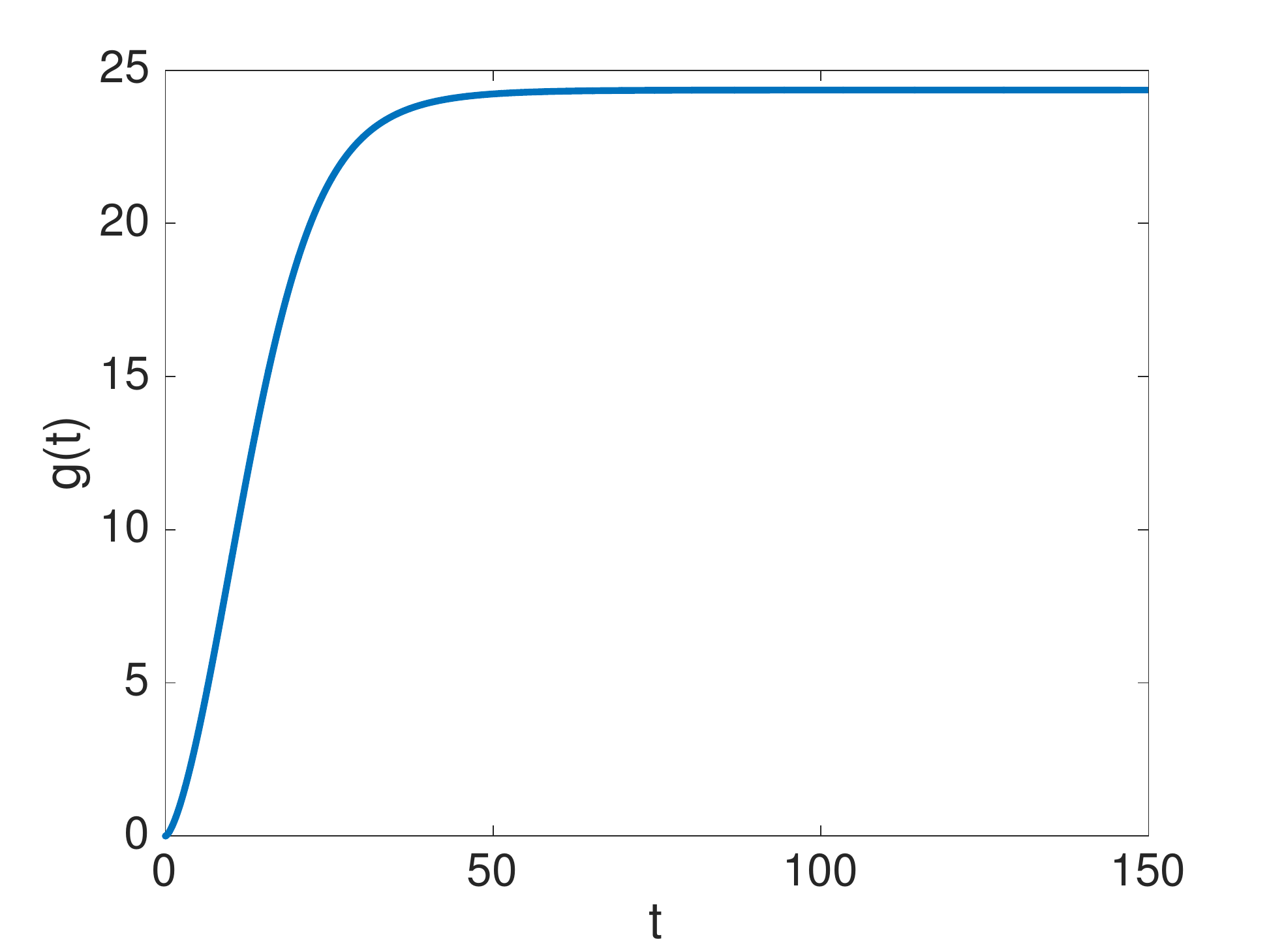}%
\label{fig:g-diffusion}}}
\centerline{\subfloat[Evolution of $\phi(\cdot,t)$ up to $t = 150$]{\includegraphics[scale=1.075]{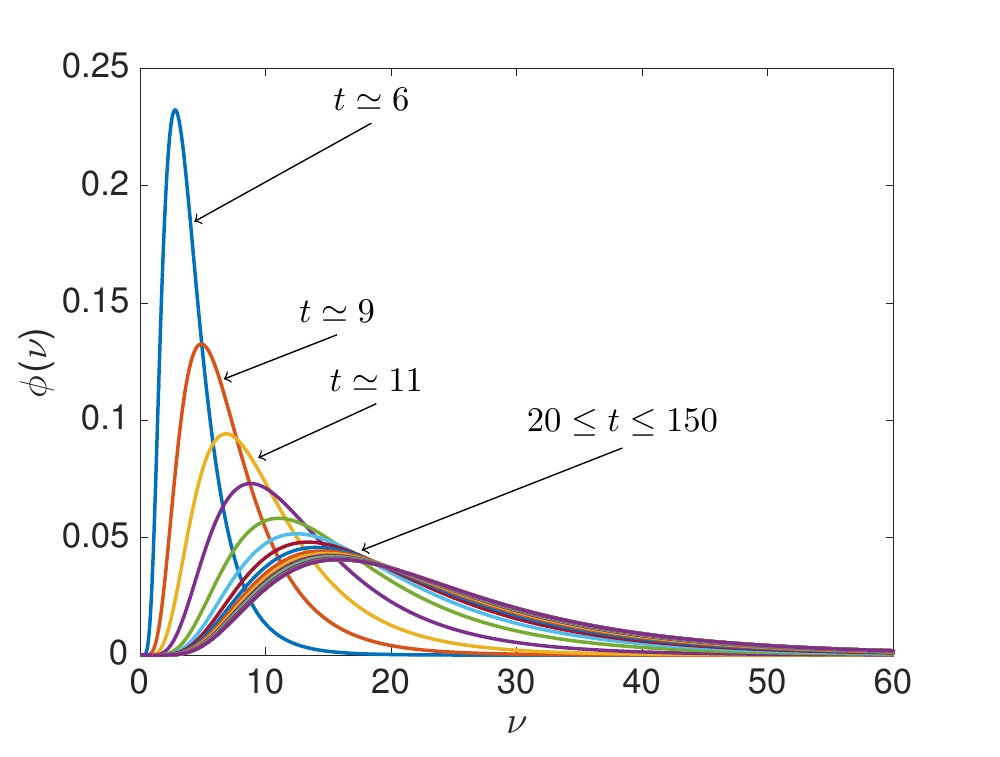}%
\label{fig:t150-diffusion}}}
\caption{Evolution of the main quantities.}
\label{fig:solution}
\efig

\subsection{Parameter study}

Figures~\ref{fig:parameter-study-rho0-solution} and~\ref{fig:parameter-study-peq-solution} show that the effect of
increasing the reference growth rate $\rho_*$ or the equilibrium phase fraction $P^\mt{eq}$ turns out to be to make the
profile flat and drift to the right more quickly. The effect of increasing the nucleation rate $\alpha$ or the power
$\gamma$ is the opposite; see Figures~\ref{fig:parameter-study-alpha-solution} and
\ref{fig:parameter-study-gamma-solution}. Increasing the diffusion coefficient $\beta_1$ or the initial standard
deviation $\sigma_0$ makes the solution flatten and shift to the left, as shown in
Figures~\ref{fig:parameter-study-beta1-solution} and~\ref{fig:parameter-study-sigma0-solution}.

\bfig[!ht] 
\centerline{\subfloat[Variation of $\rho_*$]{\includegraphics[scale=0.4]{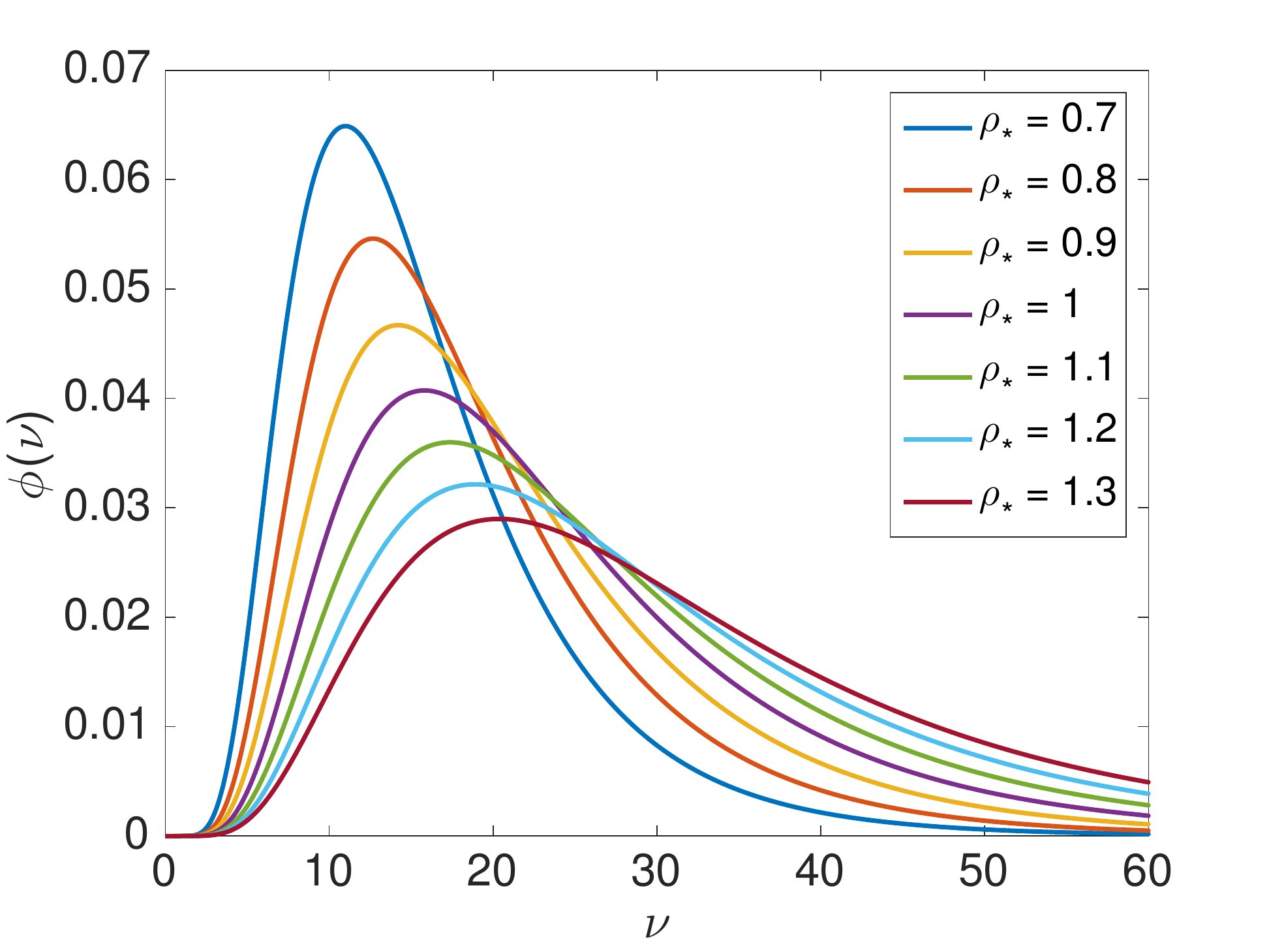}%
\label{fig:parameter-study-rho0-solution}}
\hfil
\subfloat[Variation of $P^\mt{eq}$]{\includegraphics[scale=0.4]{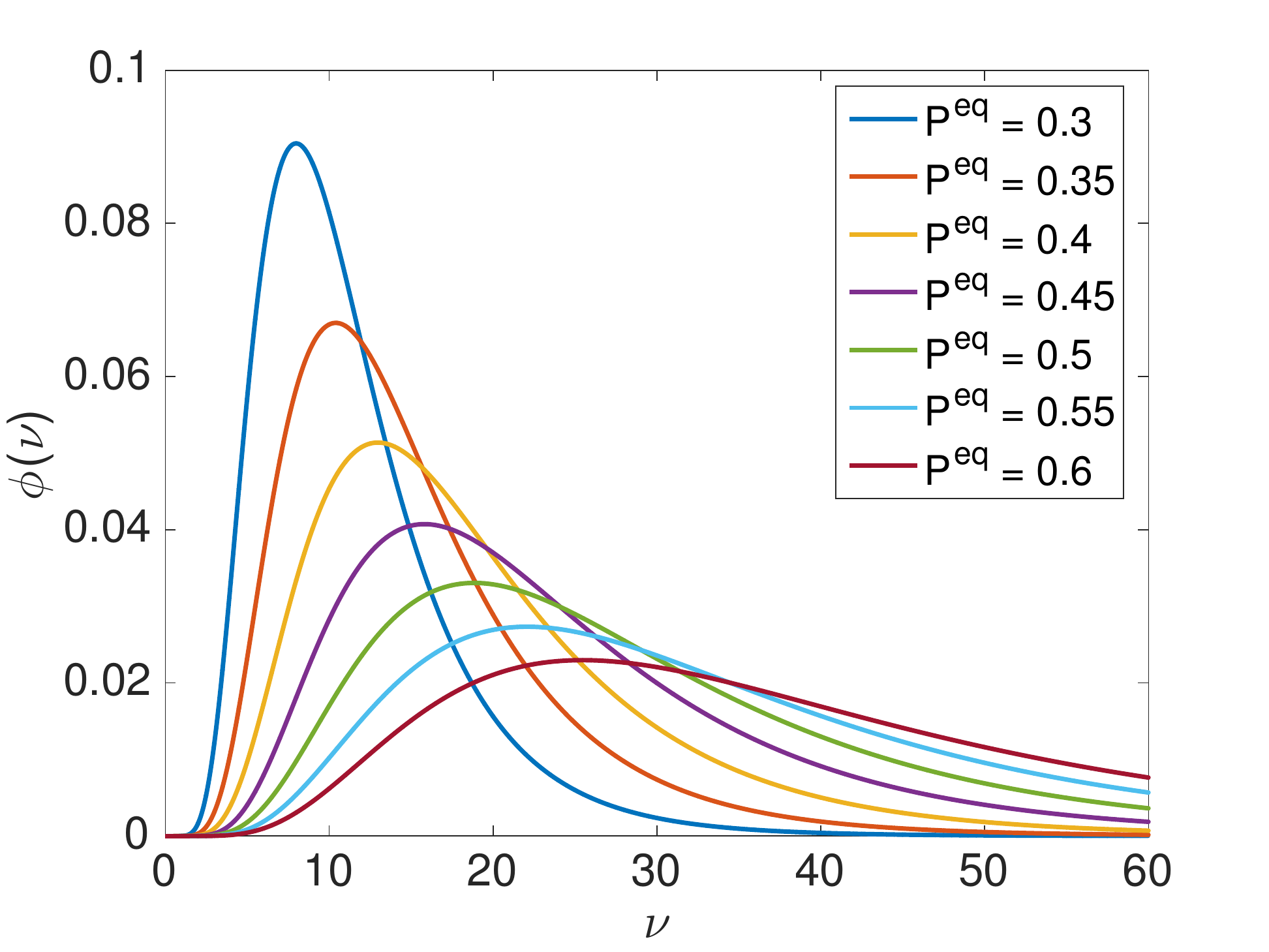}%
\label{fig:parameter-study-peq-solution}}}
\centerline{\subfloat[Variation of $\alpha$]{\includegraphics[scale=0.4]{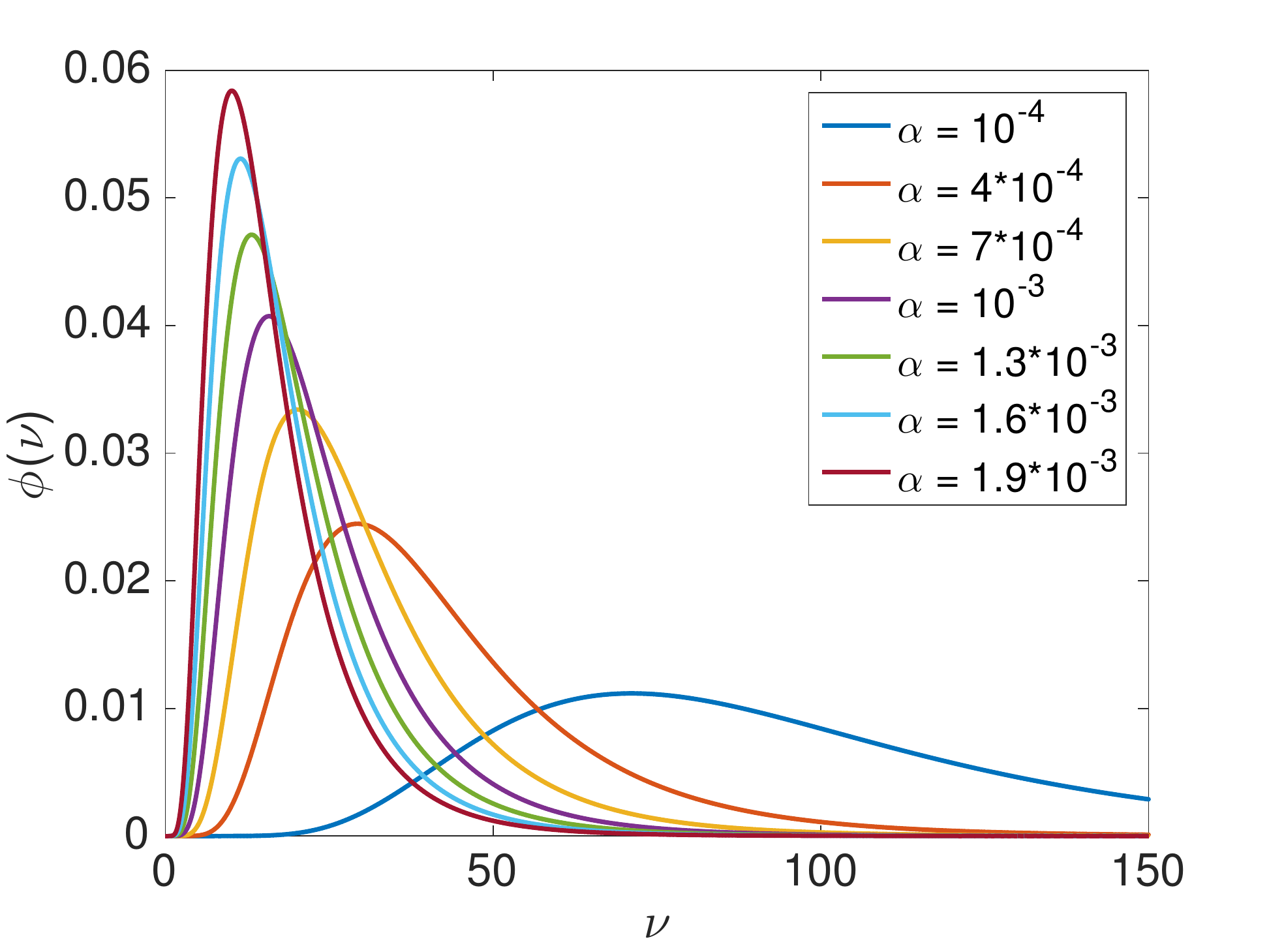}%
\label{fig:parameter-study-alpha-solution}}
\hfil
\subfloat[Variation of $\gamma$]{\includegraphics[scale=0.4]{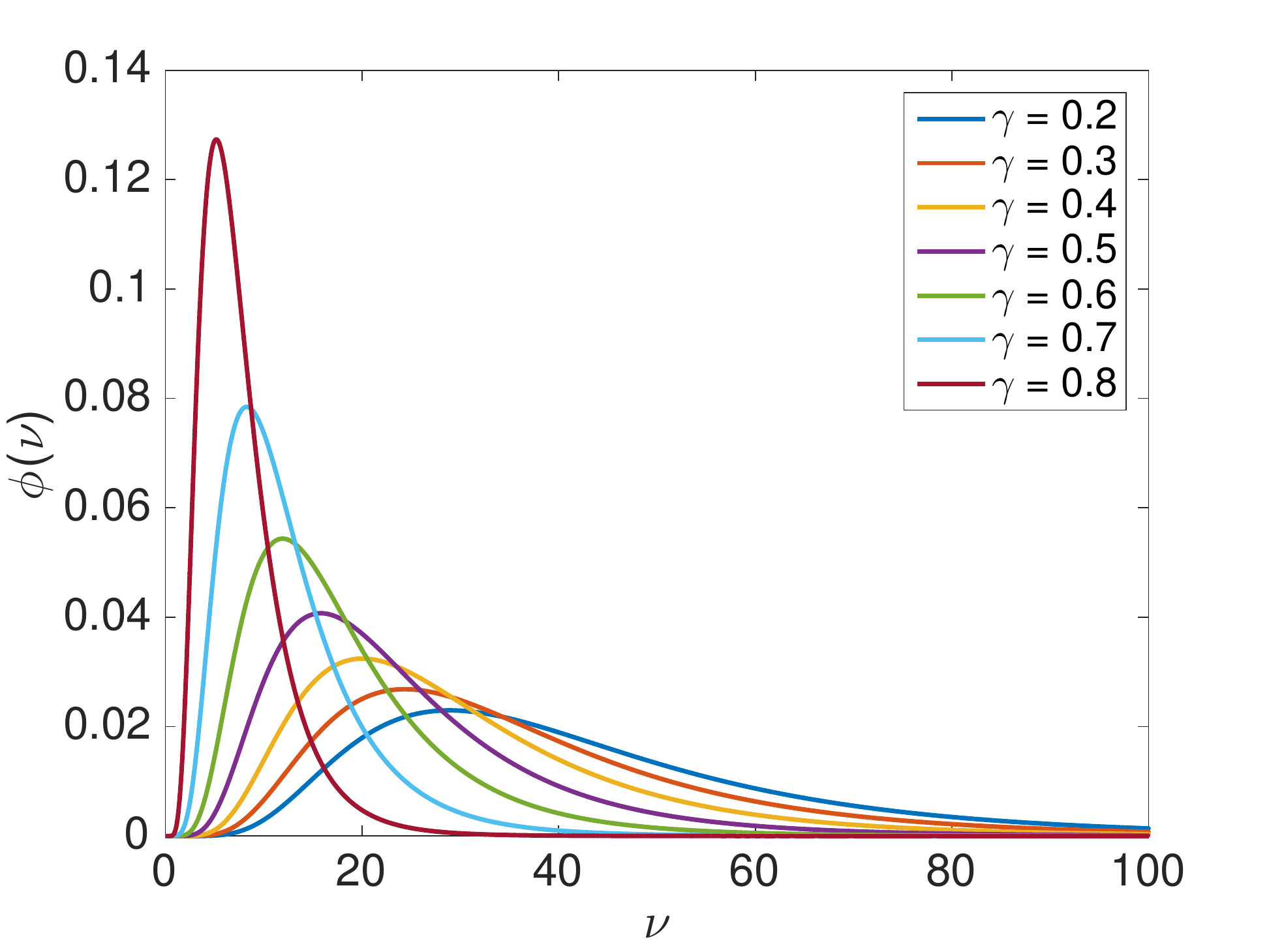}%
\label{fig:parameter-study-gamma-solution}}}
\centerline{\subfloat[Variation of $\beta_1$]{\includegraphics[scale=0.4]{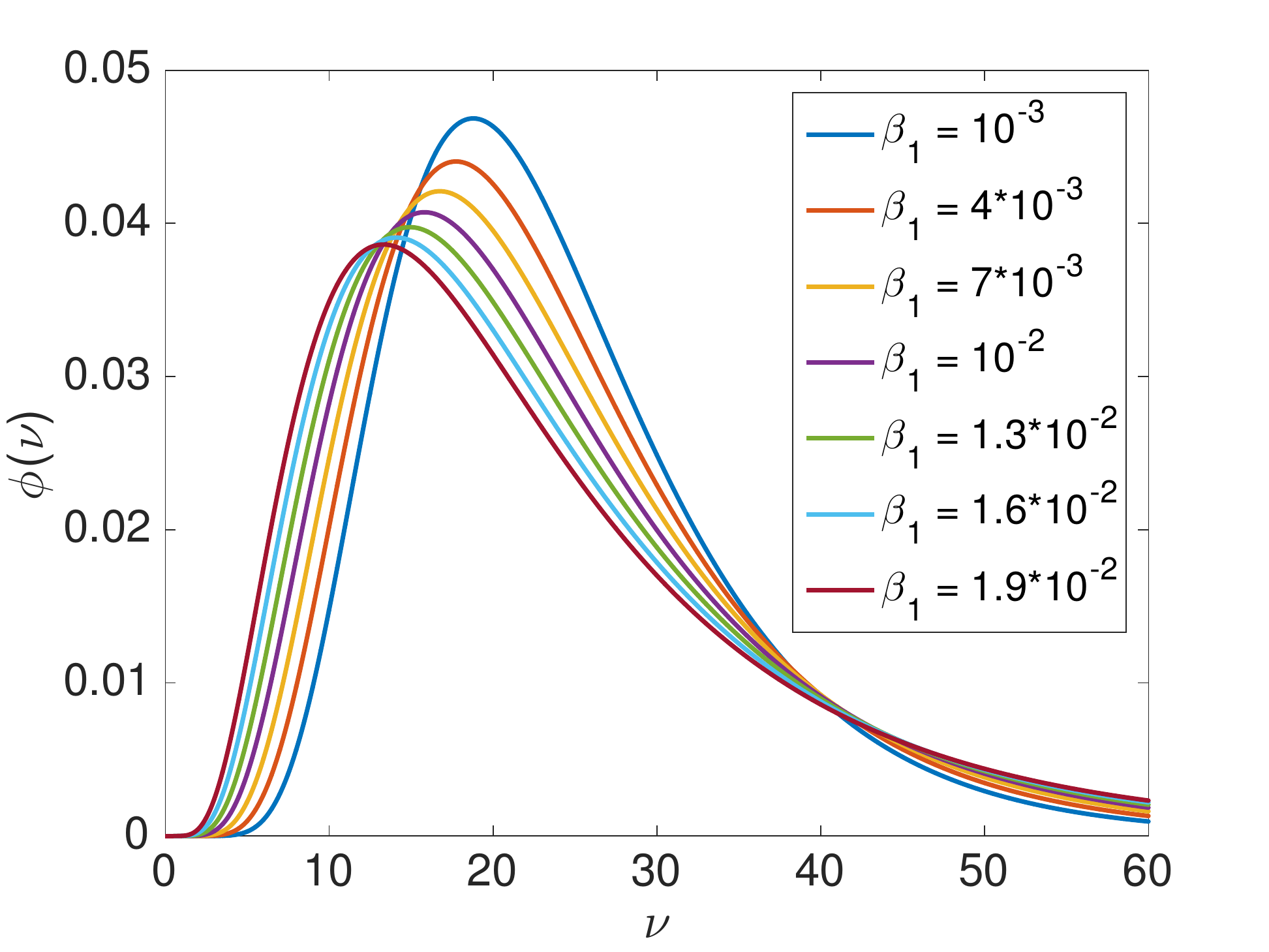}%
\label{fig:parameter-study-beta1-solution}}
\hfil
\subfloat[Variation of $\sigma_0$]{\includegraphics[scale=0.4]{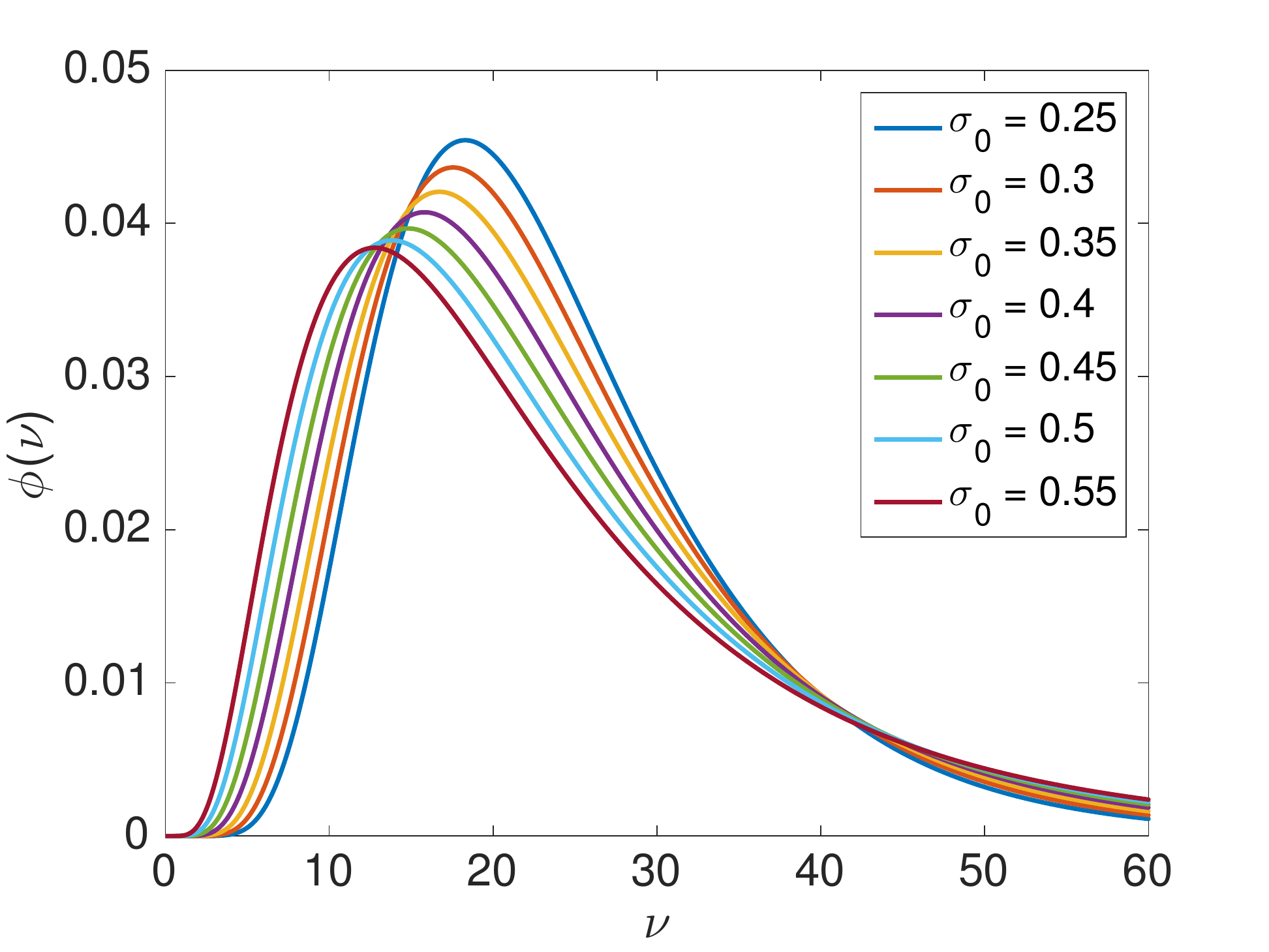}%
\label{fig:parameter-study-sigma0-solution}}}
\caption{Volume distribution at $t = 150$ for different parameter variations.}
\label{fig:parameter-study-solution}
\efig

In Figures~\ref{fig:parameter-study-rho0-solution2}, \ref{fig:parameter-study-peq-solution2} and~\ref{fig:parameter-study-alpha-solution2}, one sees that the effect of increasing $\rho_*$, $P^\mt{eq}$ or $\alpha$ is
to make the solution approach the equilibrium faster. Figure~\ref{fig:parameter-study-gamma-solution2} shows that
increasing $\gamma$ has the contrary effect.

\bfig[!ht] 
\centerline{\subfloat[Variation of $\rho_*$]{\includegraphics[scale=0.4]{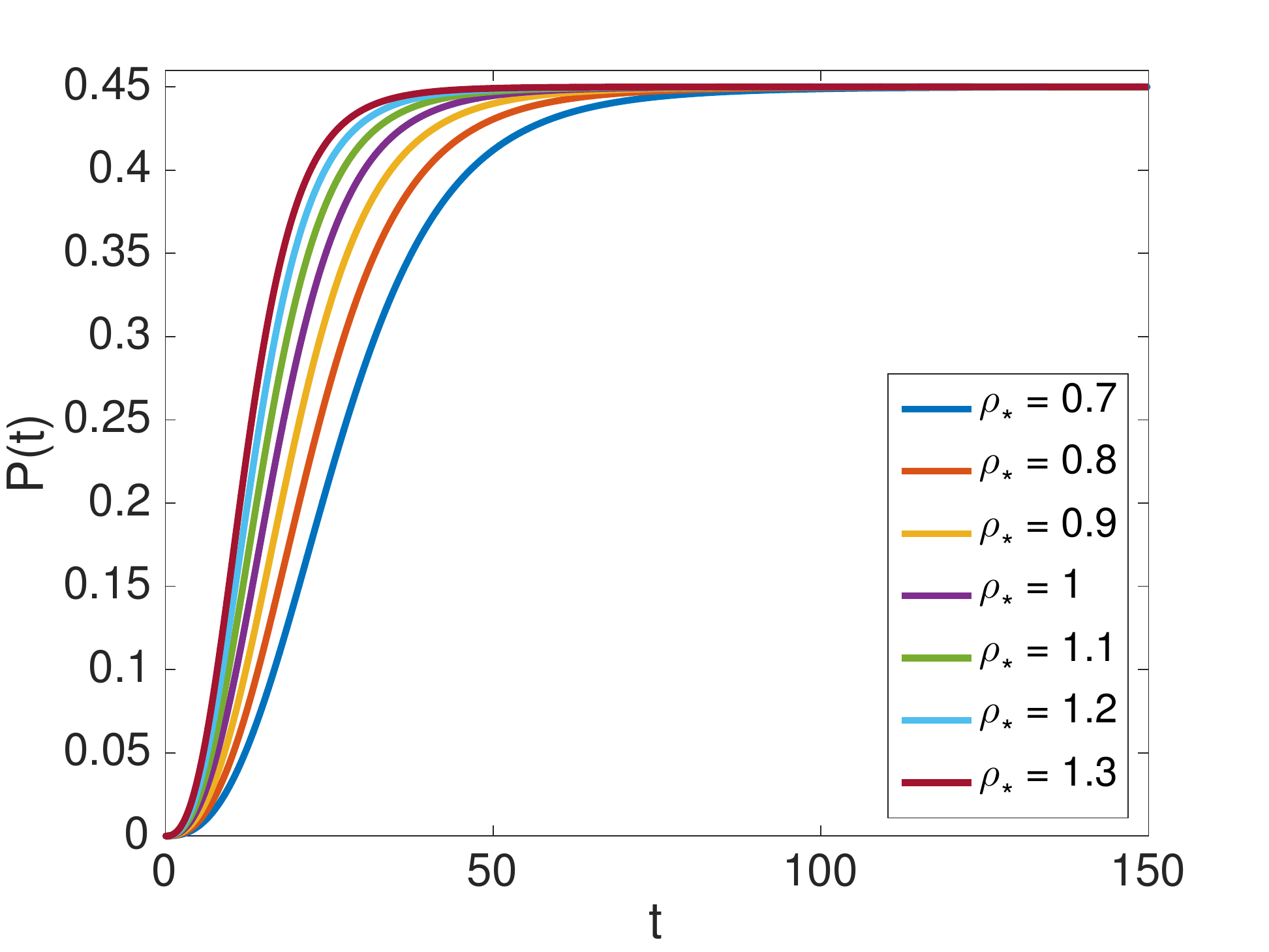}%
\label{fig:parameter-study-rho0-solution2}}
\hfil
\subfloat[Variation of $P^\mt{eq}$]{\includegraphics[scale=0.4]{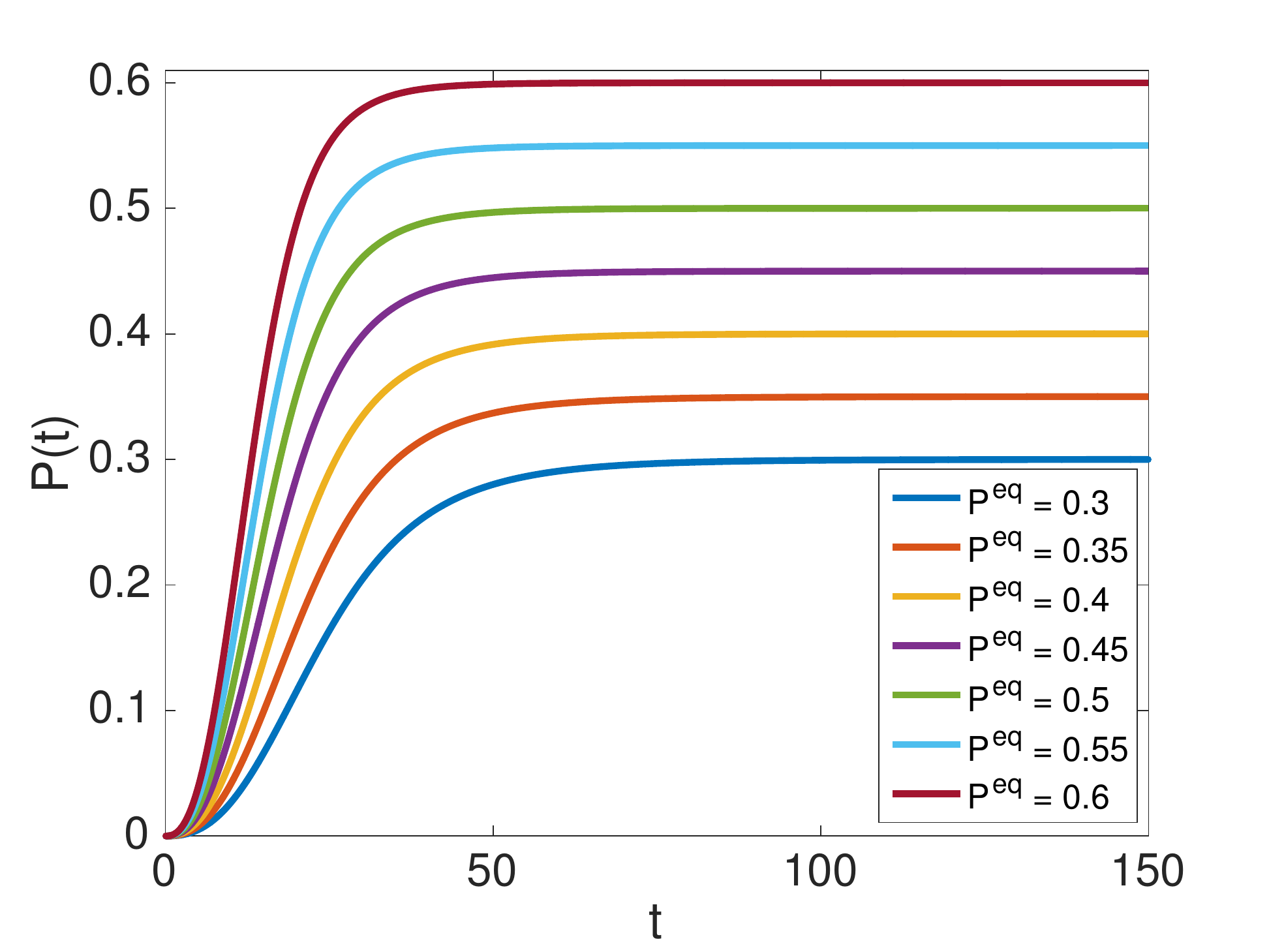}%
\label{fig:parameter-study-peq-solution2}}}
\centerline{\subfloat[Variation of $\alpha$]{\includegraphics[scale=0.4]{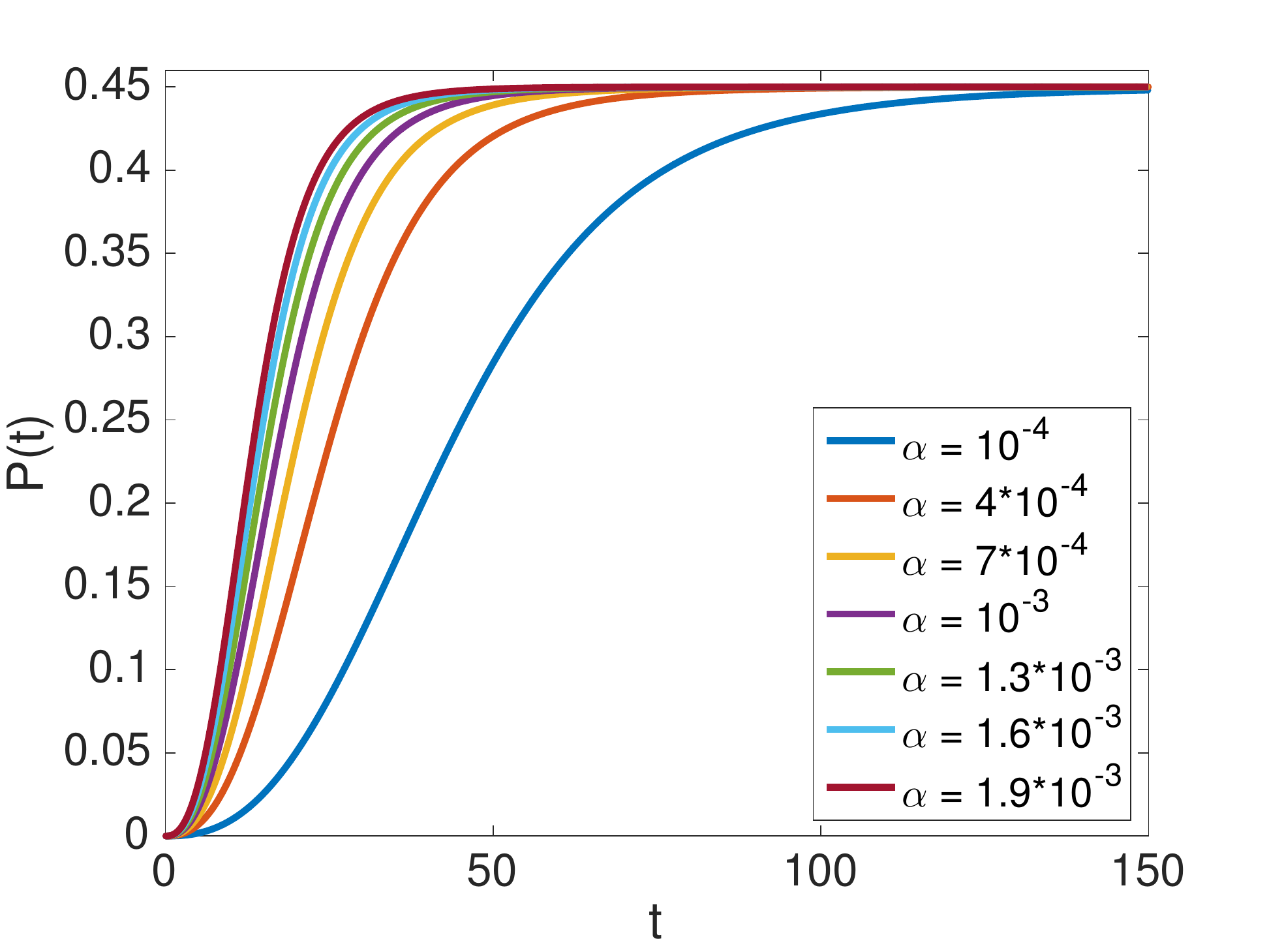}%
\label{fig:parameter-study-alpha-solution2}}
\hfil
\subfloat[Variation of $\gamma$]{\includegraphics[scale=0.4]{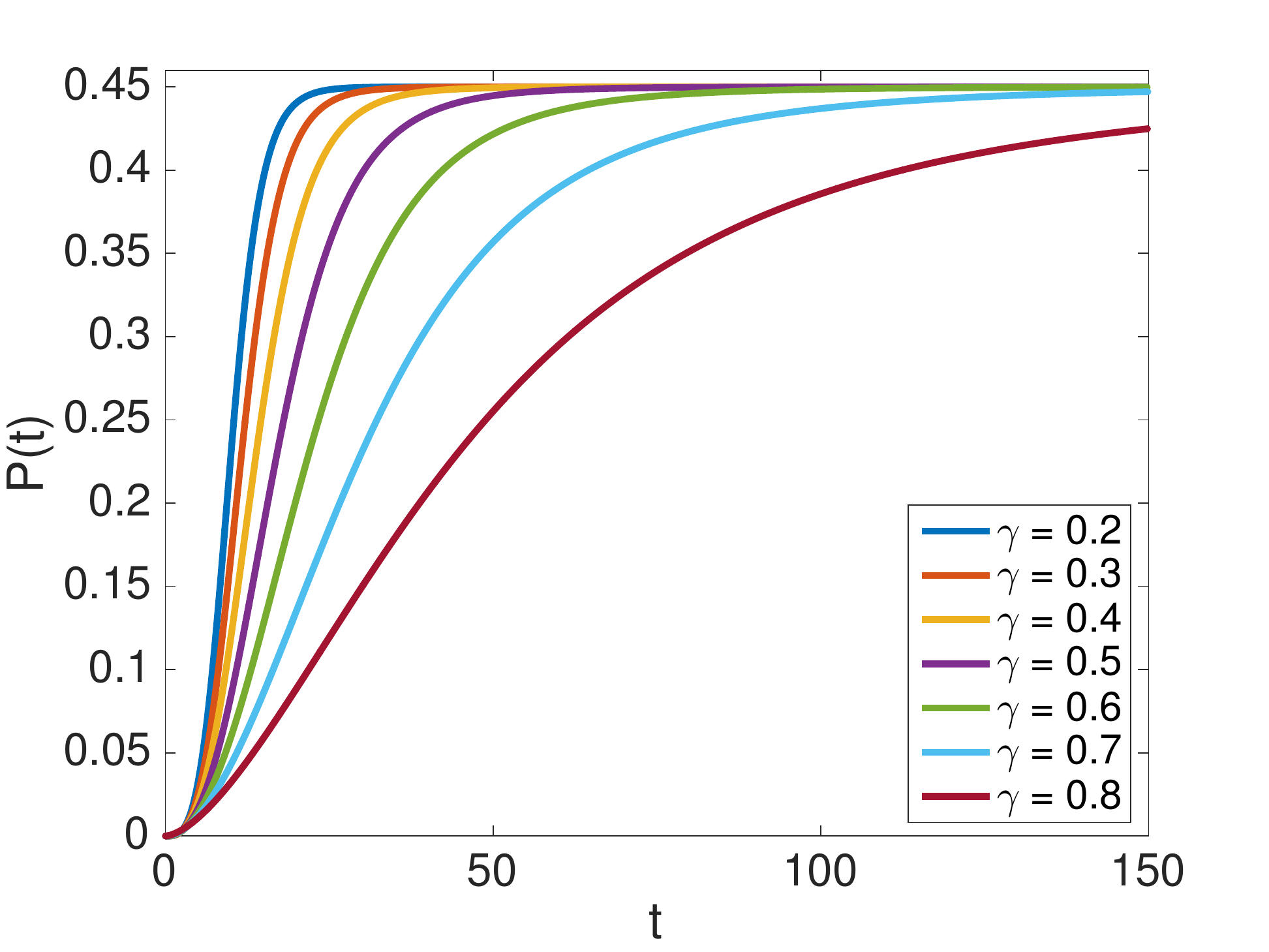}%
\label{fig:parameter-study-gamma-solution2}}}
\caption{Ferrite volume phase fraction at $t = 150$ for different parameter variations.}
\label{fig:parameter-study-solution2}
\efig

\subsection{The initial profile}\label{subsec:initial-profile}

We start here with a different initial profile from~\eqref{eq:initial}, which is instead compactly supported and therefore not log-normally distributed: 
\be\label{eq:phi0-compact}
	\phi_0(\nu) = \begin{cases} c\exp\left( \frac{-1}{k-(\nu-\nu_0)^2} \right) \quad \mbox{for all $\nu\in \left[\nu_0-\sqrt{k},\nu_0+\sqrt{k}\right]$},\\ 0 \quad \mbox{otherwise}. \end{cases}
\ee
where $k=0.1$, $\nu_0=\sqrt{k}+0.1$ and $c$ is the normalising constant. The simulation in Figure~\ref{fig:compact} shows that, qualitatively, the solution evolves into a shape very close to that in Figure~\ref{fig:t150-diffusion}, although the volume range is much larger. This supports the fact that our solution~\eqref{eq:sol1} is log-normal asymptotically in time, independently of the initial datum. The parameters used to obtain Figure~\ref{fig:solution-compact} are those in~\eqref{eq:parameters}, except for $\sigma_0$ which does not play a role in this case. 

\bfig[!ht] 
\centerline{\subfloat[Compactly supported initial profile~\eqref{eq:phi0-compact}]{\includegraphics[scale=0.4]{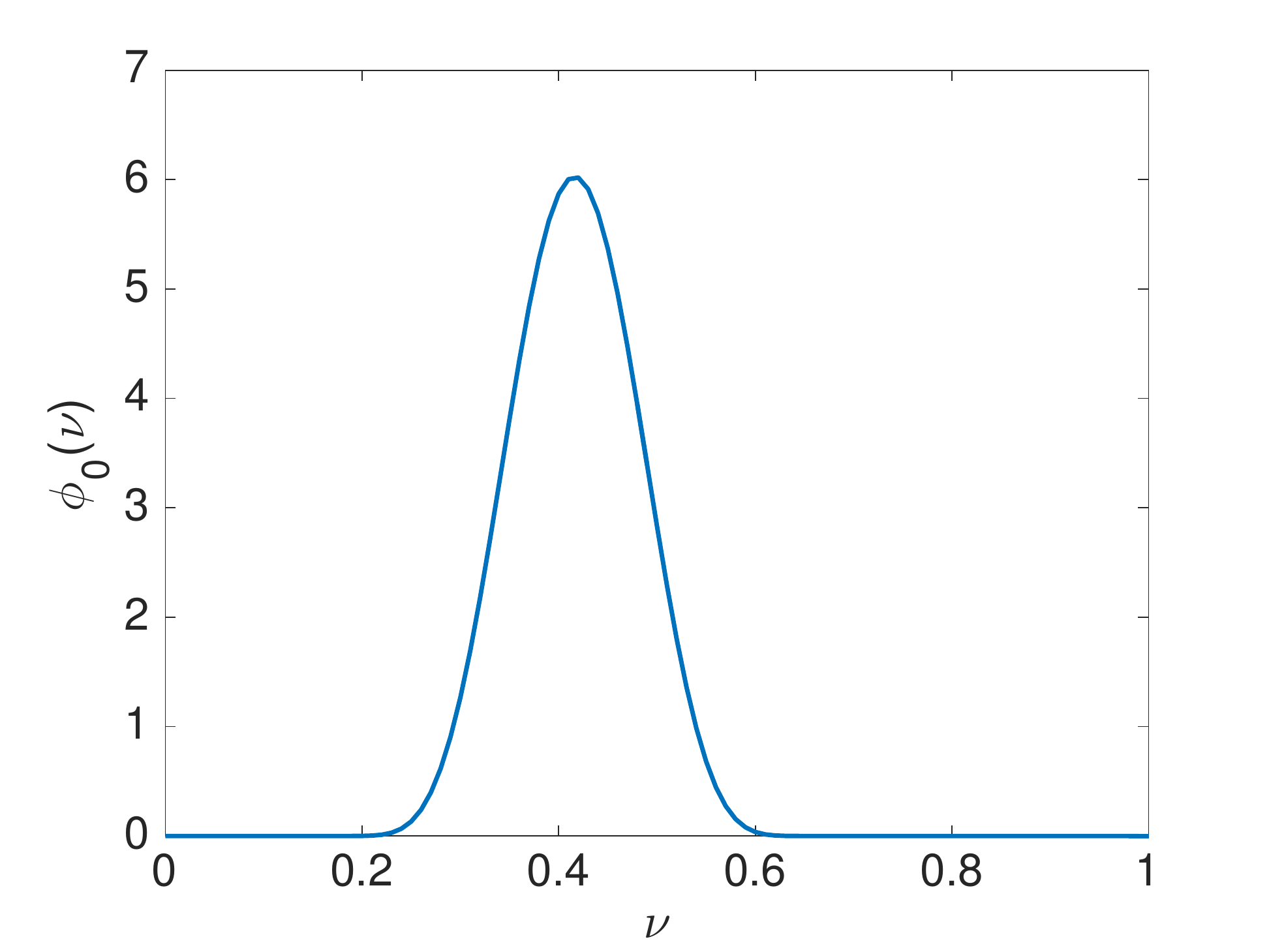}%
\label{fig:initial-compact}}
\hfil
\subfloat[Evolution of $\phi(\cdot,t)$ up to $t = 150$]{\includegraphics[scale=0.4]{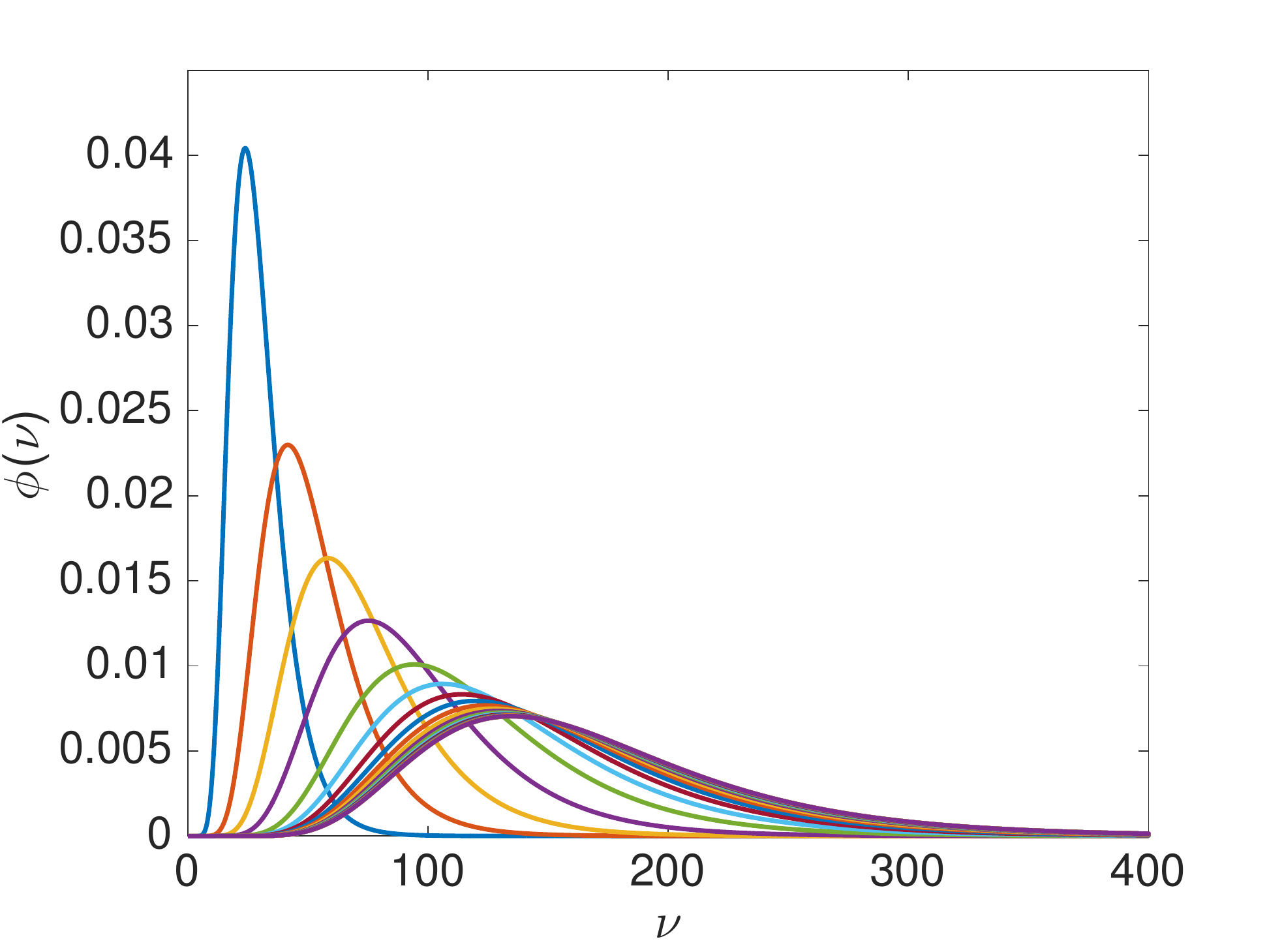}%
\label{fig:solution-compact}}}
\caption{The case of a compactly supported initial profile.}
\label{fig:compact}
\efig

\subsection{Infinite-time blow-up}\label{subsec:blowup}

We observe here the behaviour of the solution $\phi$ when the condition $f(0)=0$ is violated in $\beta(t) = f(u(t))$ in
the special case $f(u) = \beta_0 + \beta_1 u^2$ for $\beta_0 = 0.005$. Since $g$ goes to an equilibrium value as $t$
increases and $u = g'/g$ by~\eqref{eq:v}, then $u$ tends to $0$ (if $g$ does not oscillate around its equilibrium
value). Therefore $f(u(t))$ tends to $\beta_0 \neq 0$ as $t$ increases, and the Fokker-Planck equation~\eqref{eq:fp}
qualitatively becomes
\begin{equation}
  \phi_t = \beta_0(\nu^2\phi)_{\nu\nu},
\end{equation}
whose solution blows up in infinite time towards a Dirac mass at the origin, as illustrated in Figure~\ref{fig:behaviour-beta0-nonzero}. The way this solution converges to a Dirac mass is in a very weak sense; indeed, one can check that its moments do not go to $0$, but rather to a positive constant or to $+\infty$.

From Figure~\ref{fig:beta0150}, the solution first drifts to the right, until the diffusion takes over and makes the
solution drift to the left. Infinite-time blow-up occurs; see Figure~\ref{fig:beta01000}.

\bfig[!ht]
\centerline{\subfloat[Evolution of $\phi(\cdot,t)$ up to $t = 150$]{\includegraphics[scale=1.075]{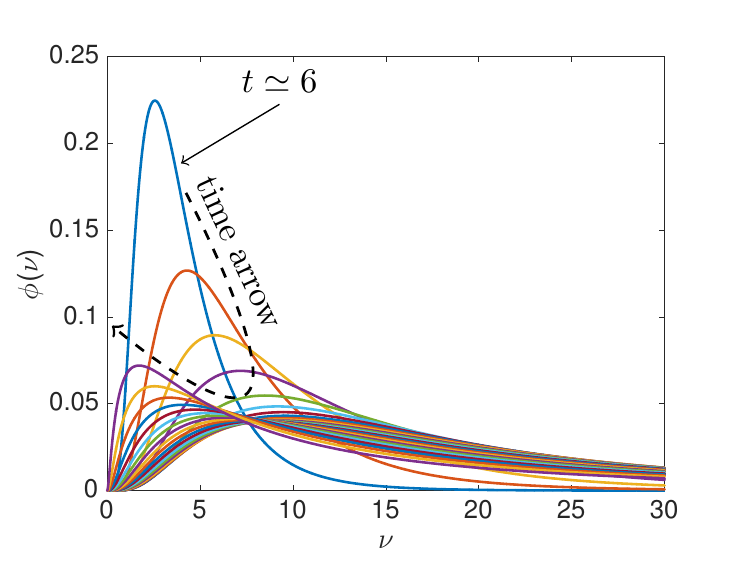}%
\label{fig:beta0150}}
\hfil
\subfloat[Evolution of $\phi(\cdot,t)$ up to $t=1000$]{\includegraphics[scale=1.075]{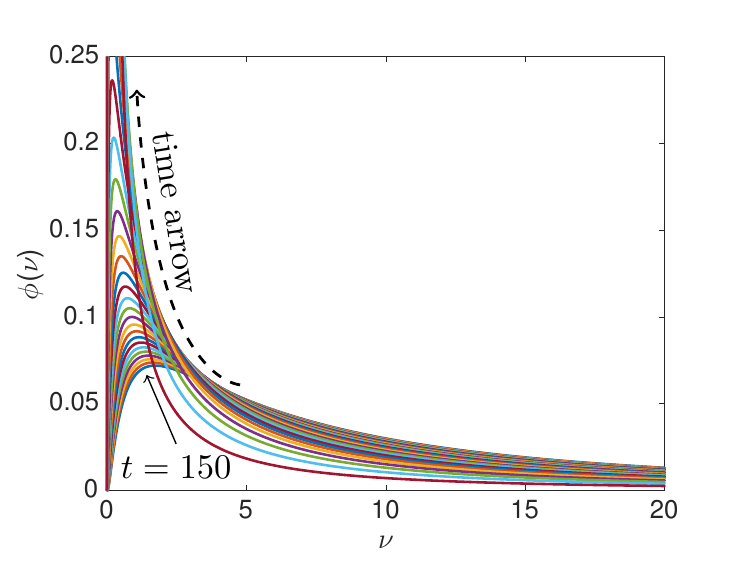}%
\label{fig:beta01000}}}
\caption{Infinite-time blow-up for $\beta_0 = 0.005$.}
\label{fig:behaviour-beta0-nonzero}
\efig

\section{Volume and area distributions: relation between model and experiments}
\label{sec:model-data}

To validate our grain-size model, the resulting volume distribution has to be related to experimental data which are
typically derived from a two-dimensional micrograph section, under the form of an area distribution. We here derive a relationship between these two distributions which can pave the
way to a quantitative validation with measurements in a forthcoming paper. We point out that the following derivation
only holds in the setting of Remark~\ref{rem:radius}; we therefore equivalently deal with three-dimensional and
two-dimensional radius distributions instead of volume and area distributions, respectively.

We follow the approach of~\cite{Huber} and consider a cylindrically shaped steel specimen $\Omega$ with base area $q$,
axially symmetric to the $z$-axis and of length $L\gg \sqrt{q/\pi}$. We want to relate the spherical grains
in $\Omega$ with their two-dimensional counterparts, that is, with the discs resulting from the intersection of the
spherical grains with the plane $\{z=0\}$. Let us fix a time $t$, and define $\chi(\eta,t)$ as the number of such
intersection discs with radius $\eta$ per unit radius per unit surface. (Note that, unlike $\psi$, $\chi$ is not
normalised by the total number of circular grains (intersection discs) in the cross-section $\Omega \cap \{z=0\}$, but
it is rather a quantity per unit surface.) Due to the boundedness of the test specimen, we may assume that the radius
of the spherical grains is bounded by some $\rmax \leq \sqrt{q/\pi}$, so that $0\leq \eta\leq\rmax$. Now let us
choose $\eta \in [0,\rmax)$ and $\de>0$ small. Then the number of circular grains in the cross-section with radii in
$[\eta,\eta+\de]$ is given by $q\int_\eta^{\eta+\de} \chi(\zeta,t) \d \zeta$. To relate this to $\psi$, we fix
$\dr > 0$ small and, for any spherical grain radius $r\in [\eta+\de,\rmax-\dr]$, we infer that the centres of the
spherical grains in the right part of the cylindrical specimen (i.e., in $\Omega \cap \{z\geq0\}$) with radii in
$[r,r+\dr]$ creating intersection discs with radii in $[\eta,\eta+\de]$ lie in an interval $[\tz,\tz+\delta]$, as shown
in Figure~\ref{huber}.

\bfig[!ht]
   \centering
\includegraphics[scale=0.5]{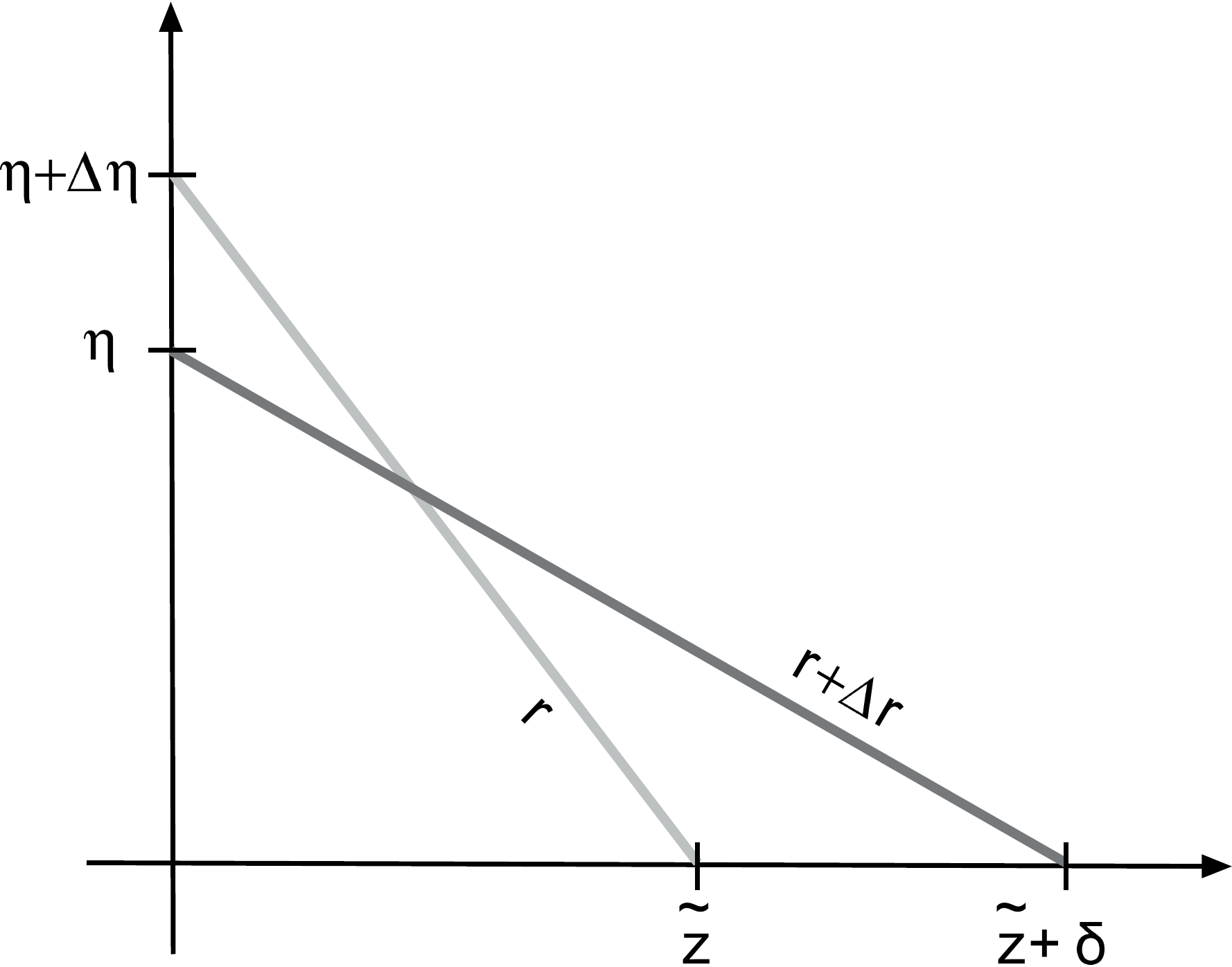}
   \caption{Position of spherical grain centres in right half of specimen.}
   \label{huber}
\efig

Note that $\delta$ depends on $\eta, \de, r$ and $\dr$; also, for $r=\eta+\de$, we have $\tz=0$. As it is immediate
from Figure~\ref{huber}, we have
\begin{equation}
  \label{delta}
  \delta = \delta(r) = \sqrt{(r+\dr)^2-\eta^2}-\sqrt{r^2-(\eta+\de)^2}.
\end{equation}
Now we choose $n\in \N$, define $\dr=(\rmax-(\eta+\de))/n$ and introduce the equi-spaced partition
\begin{equation}
  r_i=\eta+\de +i \dr, \quad 0\leq i\leq n,
\end{equation}
and accordingly $\delta_i=\delta(r_i)$; see~\eqref{delta}. A first order Taylor expansion yields
\begin{equation}
  \delta_i=c_i \eta \de +c_i r_i\dr +o(\de) + o(\dr),
\end{equation}
with $c_i:=(r_i^2-\eta^2)^{-1/2}$. Note that, by the boundedness of $r_i$, the terms $o(\de)$ and $o(\dr)$ in the above
formula are uniform in $i$. In the limit $\dr\to0$, for any $0\leq i\leq n-1$, the total number of spherical grains in
$\Omega\cap\{z\geq0\}$ with radii in $[r_i,r_{i+1}]$ contributing to circular grains in the cross-section with radii in
$[\eta,\eta+\de]$ is $N(t)q \delta_i \int_{r_i}^{r_{i+1}} \psi(r,t)\d r$, with $N(t)$ being the number of ferrite
grains per unit volume given in~\eqref{eq:N}. We sum over all transversal cylindrical pieces of $\Omega$ with volumes
$q \delta_i$, accounting also for those in the left part of the specimen $\Omega\cap\{z\leq0\}$ (hence, by symmetry,
the factor 2 in the computation below), and use the boundedness of $\psi$ and $\psi_r$ to obtain
\begin{align}
  q\int_\eta^{\eta+\de} \chi(\zeta,t)\d \zeta 
  &= 2N(t) q \sum_{i=0}^{n-1} \delta_i \int_{r_i}^{r_{i+1}} \psi(r,t)\d r + \e(\dr)\\
  &=2N(t) q \sum_{i=0}^{n-1} \delta_i\Big( \dr \psi(r_i,t) + o(\dr) \Big) + \e(\dr)\\
  &=2N(t) q \sum_{i=0}^{n-1} \Big(c_i \eta \de +c_i r_i\dr +o(\de)+o(\dr)\Big) \Big( \dr \psi(r_i,t)+  o(\dr) \Big)\\
  &\phantom{{}={}} + \e(\dr)\\
  &=2N(t) q \sum_{i=0}^{n-1} \frac{\psi(r_i,t)\eta \de \dr}{\sqrt{r_i^2 - \eta^2}} +o(\de) + \e(\dr),
\end{align}
where $\e(\dr) \to 0$ as $\dr\to0$. By letting $\dr\to0$ in the above computation, we get
\begin{equation}
  q\int_\eta^{\eta+\de} \chi(\zeta,t)\d \zeta = 2N(t) q \int_{\eta+\de}^{\rmax} \frac{\eta \de \psi(r,t)}{\sqrt{r^2-\eta^2}} \d r +o(\de).
\end{equation}
Then, dividing by $q\de$ and passing to the limit with $\de\to 0$, we finally obtain
\begin{equation}
  \label{eq:huber}
  \chi(\eta,t)=2N(t) \eta \int_\eta^{\rmax}  \frac{ \psi(r,t)}{\sqrt{r^2-\eta^2}} \d r \quad \mbox{for all $\eta\in(0,\rmax]$}.
\end{equation}
This equation relates the three-dimensional radius distribution of a given specimen to the two\-dimensional one in a
cross-section of this specimen; note that this is independent of the area $q$ of the cross-section. In
Figure~\ref{fig:huber} we give an example of comparison between $\psi$ and $\chi$ for $\rmax = 3$, according
to~\eqref{eq:huber}; there we are given a radius distribution $\psi$ which is log-normal, as well as cut off and normalised in the range $(0,\rmax]$, and then $\chi$ is computed thanks to~\eqref{eq:huber} and normalised to have mass one in $(0,\rmax]$.

\bfig[!ht]
   \centering
\includegraphics[scale=0.5]{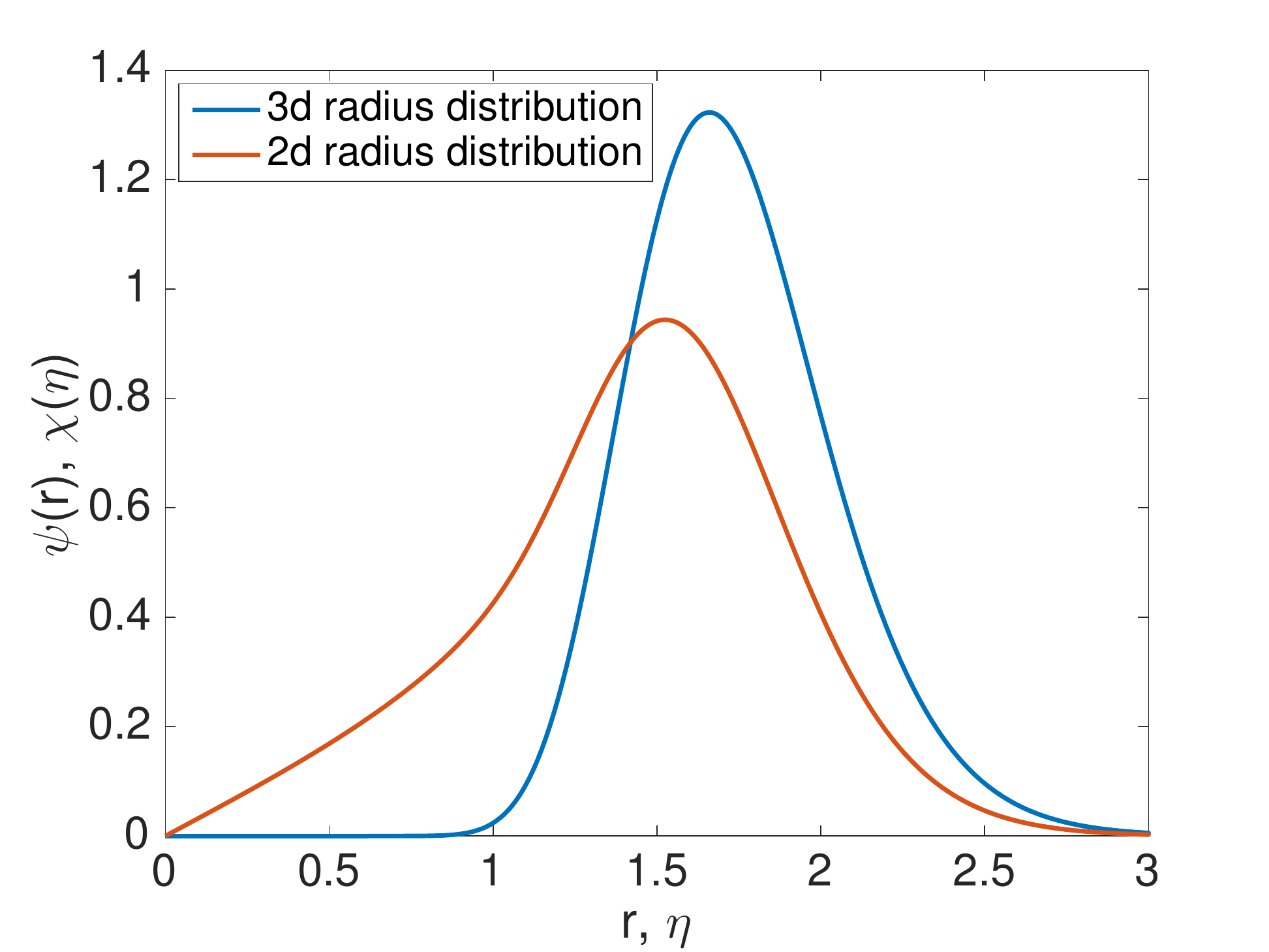}
   \caption{Three-dimensional radius distribution against two-dimensional one.} \label{fig:huber}
\efig

Let us finally point out that an easy calculation shows that the surface fraction of ferrite over the cross-section
actually coincides with its volume fraction in the specimen. Call $P_\mt{s}$ the surface fraction of ferrite, i.e., the
total surface of ferrite present on the cross-section normalised by $q$, and use formula~\eqref{eq:huber} to get, for
all $t\in[t_0,T]$,
\begin{align}
  P_\mt{s}(t) &=\pi \int_0^{\rmax} \eta^2  \chi(\eta,t) \d\eta = 2\pi N(t)\int_0^{\rmax} \eta^3   \int_\eta^{\rmax}  
                \frac{ \psi(r,t)}{\sqrt{r^2-\eta^2}} \d r \d\eta\\
              &= 2\pi N(t)\int_0^{\rmax} \psi(r,t) 
                \underbrace{\left (  \int_0^{r} \frac{\eta^3}{\sqrt{r^2-\eta^2}} \d\eta \right )}_{=2r^3/3} \d r 
                =\frac{4\pi N(t)}{3}\int_0^{\rmax}r^3\psi(r,t) \d r=P(t).
\end{align}

\section*{Acknowledgements}
D. H\"omberg gratefully acknowledges a sabbatical stay at the University of Bath where the research that led to this
work was initiated.  J. Zimmer gratefully acknowledges partial funding by the EPSRC through project EP/K027743/1, the
Leverhulme Trust, RPG-2013-261, and a Royal Society Wolfson Research Merit Award. The authors thank the anonymous reviewers for valuable suggestions. F. S. Patacchini and J. Zimmer wish to thank the Mathematisches Forschungsinstitut Oberwolfach for their support during the workshop “Applications of Optimal Transportation in the Natural Sciences” from 29 January to 4 February 2017, when progress on this paper was made.

\bibliography{grain_distribution}
\bibliographystyle{abbrv}

\end{document}